\documentclass[twocolumn,trackchanges]{aastex701}

\usepackage{amsmath,amssymb}
\providecommand{\obj}{SN~2020eyj}
\submitjournal{ApJL}
\begin{document}

%\title{Radio evolution of a Type Ia supernova reveals shock interaction with a helium-rich progenitor shell}

\title{Long-Term Radio Evolution of the Type Ia Supernova SN 2020eyj: Evidence for a Dense Helium-rich Circumstellar Shell}

% Equal contribution: Liu and Yang
% Corresponding authors: Liu, Yang, Chen, Jiang, Han
\correspondingauthor{Zheng-Wei Liu, Xiaolong Yang, Xuefei Chen, Peng Jiang, Zhanwen Han}

\author[0000-0002-7909-4171]{Zheng-Wei Liu}
\altaffiliation{These authors contributed equally.}
\affiliation{International Centre of Supernovae (ICESUN), Yunnan Key Laboratory of Supernova Research, Yunnan Observatories, Chinese Academy of Sciences (CAS), 396 Yangfangwang, Guandu District, Kunming 650216, China}
\email[show]{zwliu@ynao.ac.cn}

\author[0000-0002-4439-5580]{Xiaolong Yang}
\altaffiliation{These authors contributed equally.}
\affiliation{Shanghai Astronomical Observatory, Chinese Academy of Sciences, Shanghai 200030, China}
\affiliation{Shanghai Key Laboratory of Space Navigation and Positioning Techniques, Shanghai 200030, China}
\email[show]{yangxl@shao.ac.cn}

\author[0000-0001-8534-837X]{Huanyuan Shan}
\affiliation{Shanghai Astronomical Observatory, Chinese Academy of Sciences, Shanghai 200030, China}
\affiliation{College of Astronomy and Space Sciences, University of Chinese Academy of Sciences, CAS, Beijing, China}
\email{hyshan@shao.ac.cn}

\author[0000-0001-5284-8001]{Xuefei Chen}
\affiliation{International Centre of Supernovae (ICESUN), Yunnan Key Laboratory of Supernova Research, Yunnan Observatories, Chinese Academy of Sciences (CAS), 396 Yangfangwang, Guandu District, Kunming 650216, China}
\email[show]{cxf@ynao.ac.cn}

\author[0000-0002-5387-7952]{Peng Jiang}
\affiliation{Guizhou Radio Astronomical Observatory, Guizhou University, Guiyang, People's Republic of China}
\affiliation{National Astronomical Observatories, CAS, Beijing 100101, People's Republic of China}
\affiliation{College of Astronomy and Space Sciences, University of Chinese Academy of Sciences, CAS, Beijing, China}
\affiliation{Key Laboratory of Radio Astronomy and Technology, Chinese Academy of Sciences, Beijing, China}
\email[show]{pjiang@bao.ac.cn}

\author[0000-0001-5316-2298]{Xiangcun Meng}
\affiliation{International Centre of Supernovae (ICESUN), Yunnan Key Laboratory of Supernova Research, Yunnan Observatories, Chinese Academy of Sciences (CAS), 396 Yangfangwang, Guandu District, Kunming 650216, China}
\email{xiangcunmeng@ynao.ac.cn}

\author[0000-0002-4460-0097]{Friedrich K. R\"{o}pke}
\affiliation{Zentrum f\"ur Astronomie der Universit\"at Heidelberg, Astronomisches Rechen-Institut, M{\"o}nchhofstr.~12--14, 69120 Heidelberg, Germany}
\affiliation{Zentrum f\"ur Astronomie der Universit\"at Heidelberg, Institut f\"ur Theoretische Astrophysik, Philosophenweg 12, 69120 Heidelberg, Germany}
\affiliation{Heidelberger Institut f\"ur Theoretische Studien, Schloss-Wolfsbrunnenweg 35, 69118 Heidelberg, Germany}
\email{friedrich.roepke@h-its.org}

\author[orcid=0009-0003-4594-3715,sname='Chiba']{Ryotaro Chiba}
\affiliation{Astronomical Science Program, Graduate Institute for Advanced Studies, SOKENDAI, 2-21-1 Osawa, Mitaka, Tokyo 181-8588, Japan}
\affiliation{National Astronomical Observatory of Japan, National Institutes of Natural Sciences, 2-21-1 Osawa, Mitaka, Tokyo 181-8588, Japan}
\email[show]{ryotaro.chiba@grad.nao.ac.jp}

\author[0000-0002-2322-5232]{Jun Yang}
\affiliation{Department of Physics and Astronomy, Chalmers University of Technology, Onsala Space Observatory, 43992 Onsala, Sweden}
\email{jun.yang@chalmers.se}

\author[0000-0003-1353-9040]{Bo Zhang}
\affiliation{Shanghai Astronomical Observatory, Chinese Academy of Sciences, Shanghai 200030, China}
\affiliation{Shanghai Key Laboratory of Space Navigation and Positioning Techniques, Shanghai 200030, China}
\email{zb@shao.ac.cn}

\author[0000-0002-9728-1552]{Su Yao}
\affiliation{National Astronomical Observatories, CAS, Beijing 100101, People's Republic of China}
\affiliation{Max-Planck-Institut f\"ur Radioastronomie, Auf dem H\"ugel 69, 53121 Bonn, Germany}
\email{yaosu@bao.ac.cn}

\author{Jingxiao Luo}
\affiliation{International Centre of Supernovae (ICESUN), Yunnan Key Laboratory of Supernova Research, Yunnan Observatories, Chinese Academy of Sciences (CAS), 396 Yangfangwang, Guandu District, Kunming 650216, China}
\email{luojingxiao@ynao.ac.cn}

\author[0000-0002-8338-9677]{Philipp Podsiadlowski}
\affiliation{London Centre for Stellar Astrophysics, Vauxhall, London, UK}
\affiliation{University of Oxford, St Edmund Hall, Oxford, OX14AR, UK}
\email{podsi@hotmail.com}

\author[0000-0002-8708-0597]{Liang-Duan Liu}
\affiliation{Institute of Astrophysics, Central China Normal University, Wuhan 430079, China}
\affiliation{Key Laboratory of Quark and Lepton Physics (Central China Normal University), Ministry of Education, Wuhan 430079, China}
\email{liuld@ccnu.edu.cn}

\author[0009-0006-9211-2860]{Hai-Liang Chen}
\affiliation{International Centre of Supernovae (ICESUN), Yunnan Key Laboratory of Supernova Research, Yunnan Observatories, Chinese Academy of Sciences (CAS), 396 Yangfangwang, Guandu District, Kunming 650216, China}
\email{chenhl@ynao.ac.cn}

\author[0000-0001-6773-7830]{Shuai Zha}
\affiliation{International Centre of Supernovae (ICESUN), Yunnan Key Laboratory of Supernova Research, Yunnan Observatories, Chinese Academy of Sciences (CAS), 396 Yangfangwang, Guandu District, Kunming 650216, China}
\email{zhashuai@ynao.ac.cn}

\author[0000-0001-7369-3539]{Wu Jiang}
\affiliation{Shanghai Astronomical Observatory, Chinese Academy of Sciences, Shanghai 200030, China}
\affiliation{Shanghai Key Laboratory of Space Navigation and Positioning Techniques, Shanghai 200030, China}
\email{jiangwu@shao.ac.cn}

\author[0000-0002-9291-9254]{Rurong Chen}
\affiliation{National Astronomical Observatories, CAS, Beijing 100101, People's Republic of China}
\email{chenrr@bao.ac.cn}

\author{Jintao Luo}
\affiliation{National Time Service Center, CAS, Xi'an 710600, China}
\affiliation{Key Laboratory of Time Reference and Applications, CAS, Xi'an 710600, China}
\email{jluo@ntsc.ac.cn}

\author{De Wu}
\affiliation{National Time Service Center, CAS, Xi'an 710600, China}
\affiliation{Key Laboratory of Time Reference and Applications, CAS, Xi'an 710600, China}
\email{wude@ntsc.ac.cn}

\author[0000-0003-0721-5509]{Lang Cui}
\affiliation{Xinjiang Astronomical Observatory, CAS, 150 Science 1-Street, 830011 Urumqi, P. R. China}
\email{cuilang@xao.ac.cn}

\author[0000-0002-2577-1990]{Jiao Li}
\affiliation{National Astronomical Observatories, CAS, Beijing 100101, People's Republic of China}
\email{lijiao@bao.ac.cn}

\author[0000-0002-0584-8145]{Xiangdong Li}
\affiliation{School of Astronomy and Space Science, Nanjing University, Nanjing 210023, China}
\affiliation{Key Laboratory of Modern Astronomy and Astrophysics (Nanjing University), Ministry of Education, Nanjing 210023, China}
\email{lixd@nju.edu.cn}

\author{Ye Li}
\affiliation{Purple Mountain Observatory, Chinese Academy of Sciences, Nanjing 210023, P. R. China}
\email{yeli@pmo.ac.cn}

\author[0000-0001-9204-7778]{Zhanwen Han}
\affiliation{International Centre of Supernovae (ICESUN), Yunnan Key Laboratory of Supernova Research, Yunnan Observatories, Chinese Academy of Sciences (CAS), 396 Yangfangwang, Guandu District, Kunming 650216, China}
\email[show]{zhanwenhan@ynao.ac.cn}

\begin{abstract}
The nature of progenitor systems of Type Ia supernovae (SNe~Ia) and their explosion mechanism remain controversial. Radio observations provide a powerful probe of the interaction between SN ejecta and circumstellar material (CSM), offering important constraints on the mass-loss histories of their progenitor systems. We present extensive Jansky Very Large Array (JVLA) observations of SN~2020eyj at 1.5--10 GHz, tracing its radio emission to approximately 1300 days after discovery and revealing, for the first time, the long-term radio evolution of an SN~Ia. The radio light curves of SN~2020eyj exhibit a relatively flat evolution at early epochs, followed by a gradual long-term decline, while its radio spectra show a power-law behavior with a low-frequency turnover. These characteristics appear to be consistent with synchrotron emission associated with shock interaction. Modeling the radio light curves favors a thick circumstellar shell with an inferred mass of approximately $0.3$--$0.4\,M_\odot$ over the simple wind-like configurations. Combined with the optical spectral signatures of SN~2020eyj, our radio observations provide strong evidence for substantial helium-rich mass loss from the progenitor system prior to explosion.
The inferred CSM properties pose significant challenges to existing SN Ia progenitor models, although the ejection of a common envelope shortly before explosion may offer a plausible explanation.
\end{abstract}

\keywords{Type Ia supernovae (1728) --- Circumstellar matter (241) --- Radio transient sources (2008) --- Radio continuum emission (1340) --- White dwarf stars (1799)}

\section{Introduction}\label{sec:intro}

\begin{figure*}[ht!]
\plotone{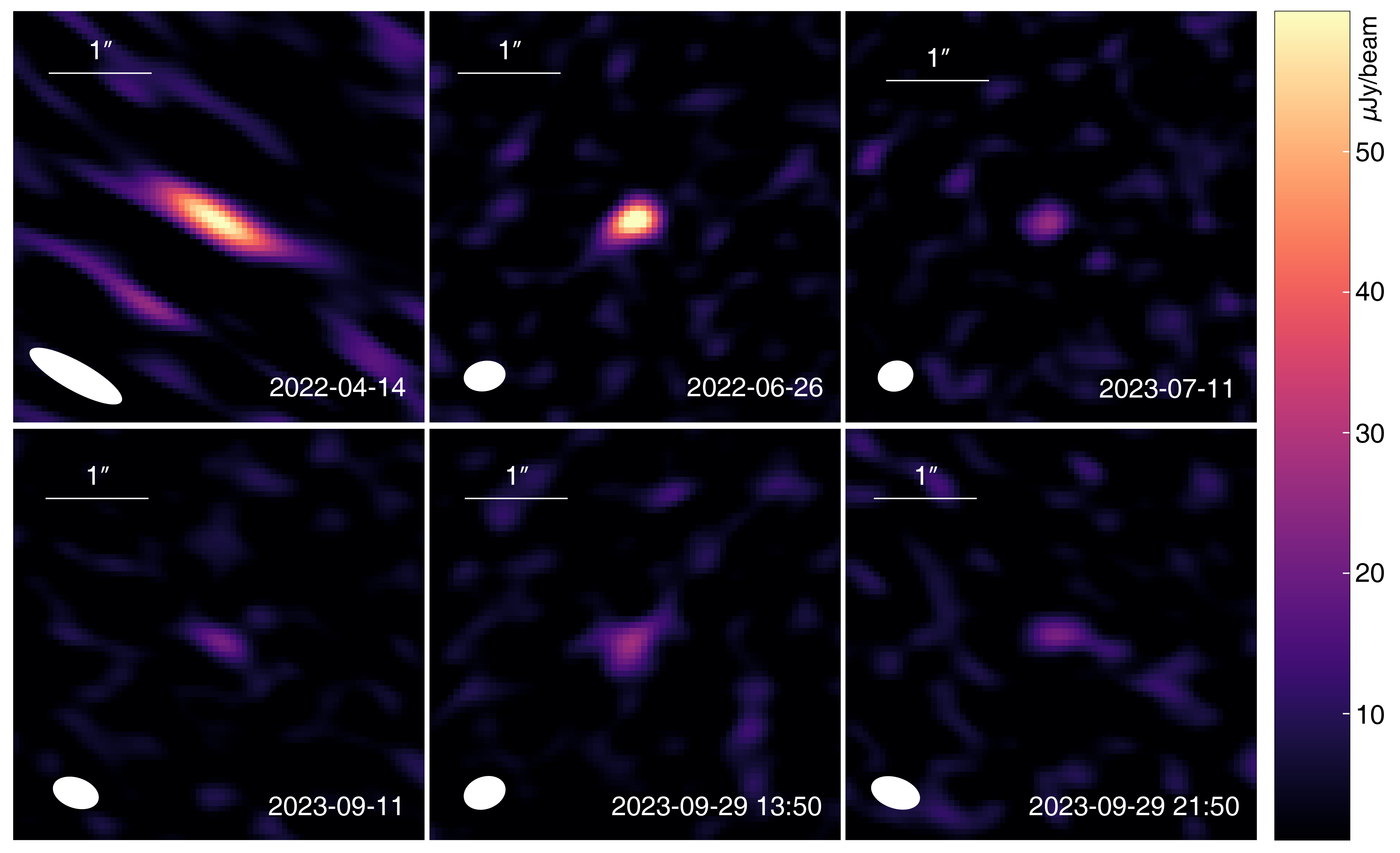}
\caption{JVLA 6\,GHz images of SN~2020eyj in six observing epochs. An overall color scale is applied to each image. The white ellipses in the bottom left corner of each sub-panel represent the full width at half-maximum (FWHM) of the synthesized beam in Table~\ref{tab:astrometry}. \label{fig:target}}
\end{figure*}

Type Ia supernovae (SNe Ia) are among the most energetic stellar explosions and play an important roles in different fields of modern astrophysics. They have been used as standardizable candles for measuring the cosmic distances, which led to the discovery of the accelerating expansion of the Universe \citep{Riess1998,Perlmutter1999}. Despite their importance, there is no a consensus on the nature of their progenitors and explosion mechanism \citep[][]{Hillebrandt2013, Maoz2014, Liu2023}. It is generally thought that SNe Ia are thermonuclear explosions of white dwarfs (WDs) in binary systems \citep{Hoyle1960}, yet the pathways leading to explosion are still debated. The two widely proposed progenitor scenarios are the single-degenerate (SD) scenario \citep[e.g.,][]{Whelan1973,Nomoto1982,Han2004} and the double-degenerate (DD) scenario \citep[e.g.,][]{Iben1984,Webbink1984,Pakmor2010,Gronow2020,Shen2025}.

After the SN explosion, the shockwaves from the explosion travel through the surrounding circumstellar material (CSM) which is formed due to pre-explosion mass loss of the progenitors, accelerating particles to relativistic velocities in amplified magnetic fields and causing them to emit strong synchrotron radiation \citep{Chevalier1982, Chevalier2006}. In the SD scenario, a significant amount of outflows is expected to form the surrounding CSM before the explosion through wind and/or binary interaction \citep{Hachisu1996, Liu2020aa}. The DD scenario is generally not expected to produce a significant amount of CSM, and no (or a weak) radio emission due to the CSM interaction is predicted \citep{Patat2007, Chomiuk2012, Maoz2014}. Therefore, one promising way to discriminate between progenitor systems is to study the radio emissions associated with SNe Ia \citep[e.g.,][]{Chomiuk2012, Chomiuk2016, Perez-Torres2014, Kundu2017, Sarbadhicary2019, Lundqvist2020}. However, before SN~2020eyj, no radio counterpart had ever been confirmed in SNe Ia \citep[e.g.][]{Chomiuk2012, Chomiuk2016, Perez-Torres2014, Kundu2017, Sarbadhicary2019, Cendes2020,Lundqvist2020, 2021ApJ...912...23H,Sand2021,Stauffer2021,Hosseinzadeh2022,Griffith2025}, even for the nearby SNe Ia SN~2011fe and SN~2014J \citep[e.g.,][]{Chomiuk2012, Lundqvist2020}.

SN~2020eyj was discovered on March 7, 2020 (UT,~${\rm MJD{=}58915.12}$), as part of the Zwicky Transient Facility (ZTF) survey in the star-forming host galaxy SDSS~J111147.15+292305.9. It was classified as a SN Ia based on a low-resolution spectrum obtained about 25 days after the first detection \citep{Dahiwale2020, Kool2023}. Given the spectroscopic redshift of $z{=}0.0297{\pm}0.0001$, and assuming a standard $\Lambda$CDM cosmology with $H_0\,{=}\,71\,\mathrm{km\,s^{-1}\,Mpc^{-1}}$, $\Omega_{\rm m}\,{=}\,0.27$, and $\Omega_{\Lambda}{=}0.73$, we calculate the distance to SN~2020eyj as $128.34\pm0.44\,\mathrm{Mpc}$. The optical light curves and spectra of SN~2020eyj exhibit typical features consistent with those of SNe~Ia-CSM systems \citep{Kool2023, Hamuy2003, Silverman2013}. In particular, it shows narrow emission lines (however, they exhibit helium lines rather than hydrogen lines) in the optical spectra, indicating a He-rich CSM. SN~2020eyj is the first SN~Ia-CSM event (to our knowledge) that exhibits clear He emission features in the spectra \citep{Simon2009, Patat2007, Sternberg2011, Kool2023}. The light curve of SN~2020eyj behaves as a typical SN Ia in the first 50 days after the detection. Thereafter, it shows significant deviations from the standard decline shape, exhibiting a plateau that lasts about $200\,\mathrm{days}$. This implies a strong CSM interaction with an inner boundary to the He-rich CSM at approximately $4\times10^{15}\,\mathrm{cm}$, for a typical ejecta expansion velocity \citep{Kool2023}.  Interestingly, two electronic Multi-Element Radio Linked Interferometer Network ($e$-MERLIN) observations identified a radio source at the position of SN~2020eyj, making SN~2020eyj as the first SN Ia detected at radio wavelengths.

In this work, we performed long-term radio observations of SN~2020eyj with the Karl~G. Jansky Very Large Array (JVLA) to obtain a time-variable radio light curve of SN~2020eyj. It aims to establish the link between previously reported radio detections with $e$-MERLIN and the actual supernova beyond doubt, and therefore place better constraints on the nature of progenitor system of SN~2020eyj. The introduction of JVLA observations and data reduction are introduced in Section~\ref{sec:observation}. In Section~\ref{sec:csm}, we present a detailed comparison between results of theoretical modelling and new JVLA observations. Some discussion of potential progenitor models of SN~2020eyj is given in Section~\ref{sec:discussion}. The main conclusions are summarized in Section~\ref{sec:summary}.

\section{VLA Observations}\label{sec:observation}
We performed radio observations of SN~2020eyj with the Karl~G. Jansky Very Large Array (JVLA), which comprises twenty-seven 25-meter radio telescopes. Radio observations were carried out at six different epochs, corresponding to days $+768$, $+780$, $+841$, $+1221$, $+1283$, and $+1301$ since the discovery, and in four different frequency bands at 1.5, 3, 6, and 10 GHz (see also Appendix~\ref{app:obs}; \citealt{Yang2023}). In Figure~\ref{fig:target}, we show the 6\,GHz images obtained from the position of SN~2020eyj, and a radio source is clearly detected in all observing epochs with signal-to-noise ratios above $\sim$5 (Table~\ref{tab:jvla}). This marks the first instance in which radio emission has been confirmed from a SN Ia up to 1301 days post-explosion, allowing us to compile an unprecedentedly detailed radio light curve of a SN Ia.

\begin{deluxetable*}{cccccccc}
\setlength{\tabcolsep}{12pt}
\tablecaption{JVLA observations of \obj. \label{tab:jvla}}
\tablehead{
\colhead{Date} & \colhead{MJD} & \colhead{$\nu$} & \colhead{$\Delta\nu$} & \colhead{TOS}   & \colhead{$f_p$} & \colhead{$\alpha$} & \colhead{Proj.ID} \\
\colhead{(UT)} & \colhead{}    & \colhead{(GHz)} & \colhead{(GHz)}       & \colhead{(min)} & \colhead{($\mu$Jy\,beam$^{-1}$)} & \colhead{} & \colhead{}
}
\colnumbers
\startdata
2022-04-14 &$ 59683.44   $&$   6      $&$        4           $&$      24.4    $&$  61.67\pm6.2        $&$     -1.591\pm0.387           $&      22A-472         \\
2022-06-26 &$ 59756.80   $&$   6      $&$        4           $&$      36.6    $&$  73.60\pm4.6        $&$     -1.134\pm0.375           $&      22A-497         \\
2023-07-11 &$ 60136.81   $&$   6      $&$        4           $&$      35.9    $&$  29.04\pm4.2        $&$     -1.016\pm0.451           $&      23A-425         \\
2023-09-11 &$ 60198.95   $&$   6      $&$        4           $&$      35.9    $&$  27.51\pm4.4        $&$     -1.033\pm0.499           $&      23B-299A        \\
2023-09-29 &$ 60216.57   $&$   6      $&$        4           $&$      48.9    $&$  25.88\pm3.7        $&$     -1.334\pm0.589           $&      23B-299C        \\
2023-09-29 &$ 60216.90   $&$   6      $&$        4           $&$      35.9    $&$  22.57\pm4.6        $&$     -1.390\pm0.745           $&      23B-299D        \\
\hline
2022-06-26 &$ 59756.80   $&$   3      $&$        2           $&$      58.3    $&$  96.43\pm7.7        $&$     -0.523\pm0.561           $&      22A-497         \\
2023-09-11 &$ 60198.99   $&$   3      $&$        2           $&$      37.9    $&$  38.78\pm7.1        $&$                              $&      23B-299B        \\   
\hline
2022-04-26 &$ 59695.97   $&$   1.52   $&$        1           $&$      52.0    $&$  82.73\pm17.6       $&$      0.992_{-1.058}^{+0.927} $&      22A-472         \\
2022-06-26 &$ 59756.80   $&$   10     $&$        4           $&$      50.1    $&$  37.93\pm3.5        $&$     -0.438\pm0.541           $&      22A-497         \\
\enddata
\tablecomments{Column 1: observing epoch; Column 2: MJD; Column 3: Frequency; Column 4: Bandwidth; Column 5: time-on-source; Column 6: peak flux density with $1\sigma$ rms noise uncertainty; Column 7: in-band radio spectral index $\alpha$ ($f_p\propto\nu^{\alpha}$); Column 8: project id.}
\end{deluxetable*}

All observations used the most extended (see Appendix~\ref{app:obs}) configuration, with J1125$+$2610 as the phase calibrator and 3C~48 (22A-472, 22A-497) or 3C~286 (23A-425, 23B-299) as the flux-density scale calibrator (Perley--Butler 2017 scale). Data were calibrated manually in CASA v6.4.1.12, imaged with \textsc{tclean} (MT-MFS, $nterms=2$, Briggs $robust=0.0$, wide-field $w$-projection/facetting), and self-calibrated using three bright in-beam sources. A relative amplitude-gain correction referenced to the first epoch (2022 April 14) was derived from these in-beam sources and applied to the target. In-band spectral indices were measured with DIFMAP \citep{1997ASPC..125...77S} by fitting a point-source model. Full details are given in Appendix~\ref{app:obs}; the resulting flux densities and spectral indices are listed in Table~\ref{tab:jvla}.

\subsection{Light curves}
The radio source shows a point-like structure in JVLA observations, where the best resolution of $0.182\times0.317$\,arcsec obtained at 10\,GHz, indicating the physical size of the radio source $<186$\,parsec. The 6\,GHz radio flux density declines from $61.67\pm6.2\,\mathrm{{\mu}Jy\,beam^{-1}}$ on April 14, 2022, to $22.57\pm4.6\,\mathrm{{\mu}Jy\,beam^{-1}}$ on September 29, 2023. In addition, the radio flux density decline was confirmed at 3\,GHz, with the peak flux density of $96.43\pm7.7\,\mathrm{{\mu}Jy\,beam^{-1}}$ on June 26, 2022, declining to $38.78\pm7.1\,\mathrm{{\mu}Jy\,beam^{-1}}$ on September 11, 2023. Involving the two pilot detections at 5\,GHz with the enhanced Multi-Element Remotely Linked Interferometer Network ($e$-MERLIN) \citep{Kool2023}, particularly the first detection on November 20, 2021, the radio light curves of SN~2020eyj show an overall declining trend from day $623$ to day $1301$ post-discovery (See Table~\ref{tab:jvla}). This is the first-ever radio light curve and spectra from an SN Ia. 

Besides the overall declining trend of the radio light curve, the most prominent feature of the radio light curve is that the 6\,GHz radio flux density remained steady from $623$ to $841$ days after discovery (See Table~\ref{tab:jvla}). Similar to the rising phase typically seen in Type II SNe \citep[e.g.,][]{2012ApJ...755..110C, 2022ApJ...934..186N}. Comparing the flux density of $61.77\pm20\,\mathrm{{\mu}Jy\,beam^{-1}}$ on November 20, 2021, the radio emission halt at $61.67\pm6.2\,\mathrm{{\mu}Jy\,beam^{-1}}$ and slightly rise to $73.60\pm4.6\,\mathrm{{\mu}Jy\,beam^{-1}}$ on April 14, 2022, and June 26, 2022, respectively. In addition to obtaining the absolute flux density scale from 3C~286 and 3C~48, and bootstrapping it to a nearby calibrator J1125+2610 and target, we performed a relative amplitude gain correction using three in-beam calibrators (within $10^\prime$ field of the target, see Appendix~\ref{app:obs}). This procedure removes calibration errors and scales flux density in this monitoring campaign. Notably, this feature is not decayed without the calibration because only the data on June 26, 2022, received relative amplitude gain correction here. For comparison, the 6\,GHz radio flux density before relative amplitude gain correction, on that date, was $70.60\pm5.8\,\mathrm{{\mu}Jy\,beam^{-1}}$.

\begin{figure}
\plotone{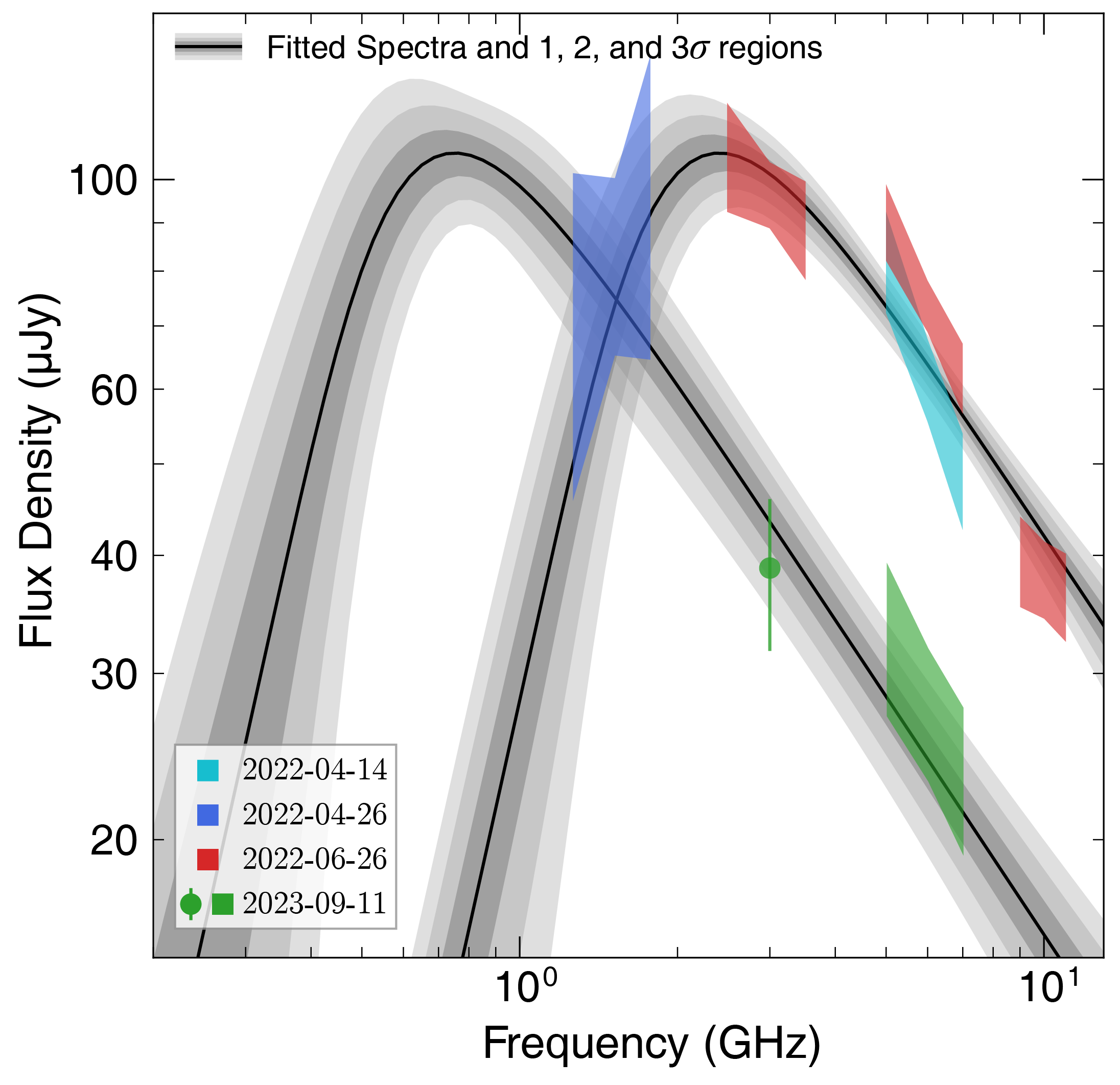}
\caption{Multi-epoch radio spectra of \obj. The peak radio flux densities are shown, where the radio flux densities extrapolate according to the flux density uncertainties and in-band spectral indexes (and their uncertainties as well). Here we only extend along half of each bandwidth. The black lines and gray belts are the best-fit spectra and the corresponding 1, 2, and 3$\sigma$ uncertainties.}
\label{fig:sp}
\end{figure}

\begin{figure*}
\plotone{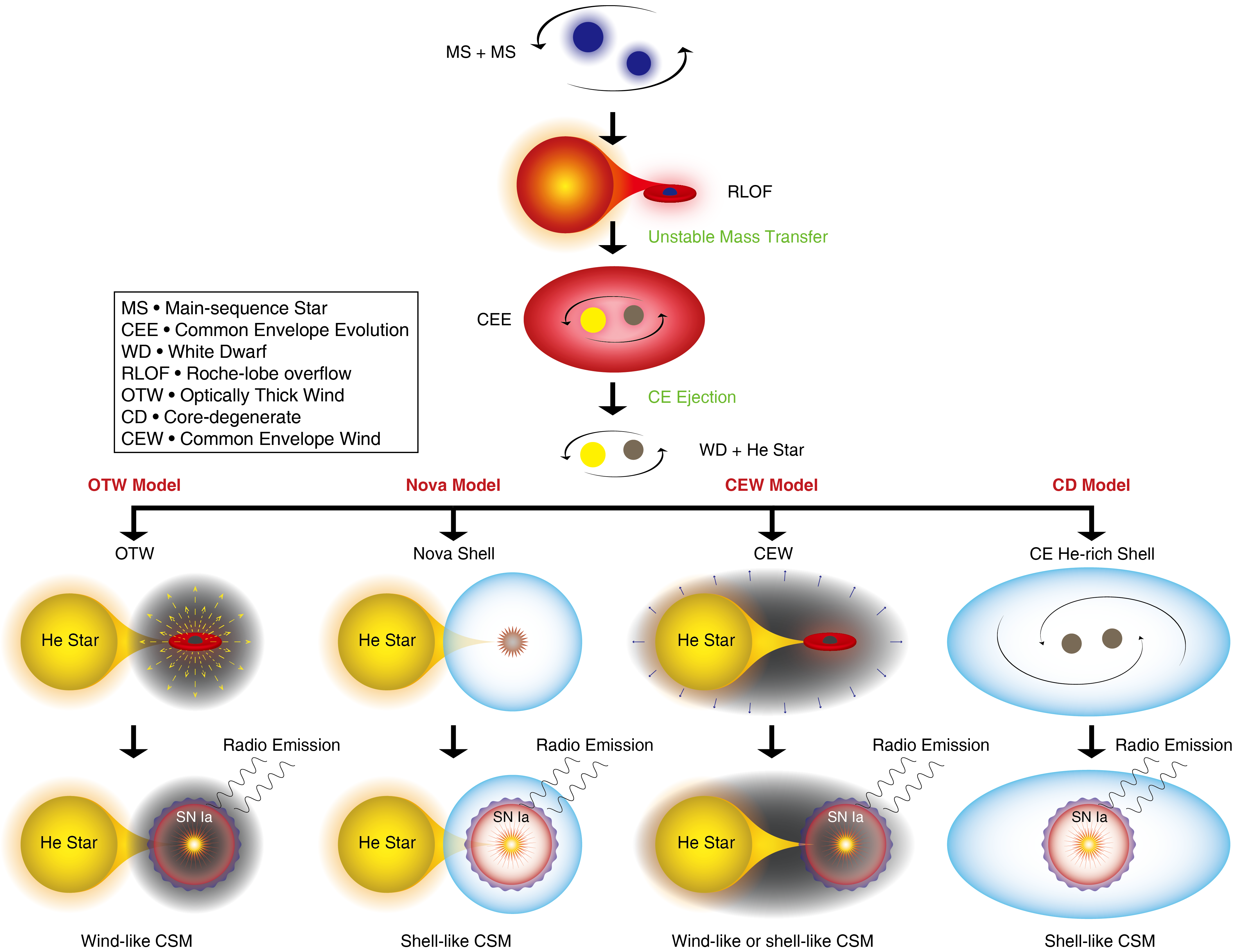}
\caption{Schematic diagram of evolutionary channels for possible progenitor models of \obj. There are four channels that can form helium-rich CSM, which is required for explaining radio emission observed in \obj, corresponding to either wind-like or shell-like structures. From left to right, it corresponds to wind-like CSM structures from the optically thick wind model (the first column branch) and the common envelope wind model (the third column branch), and shell-like structures from the nova model (the second column branch), the common envelope wind model (the third column branch) and the core-degenerate model(the fourth column branch). \label{fig:channel}}
\end{figure*}

\subsection{Radio spectra}
We also obtained the radio spectra of SN~2020eyj and correspondingly captured its spectral evolution, for the first time for a SN Ia. In Figure~\ref{fig:sp}, we show two sets of radio spectra spanning from 1 to 10\,GHz obtained from different epochs. The radio spectra of SN~2020eyj show a steep power-law decline along frequencies from 2 to 10\,GHz, with a slope of $\alpha=-0.84^{+0.11}_{-0.12}$, where flux densities are defined as $S_\nu\propto\nu^{+\alpha}$. This indicates synchrotron radio emission and corresponds to a power-law energy distribution of relativistic electrons. Interestingly, the $1.52\,\mathrm{GHz}$ flux density deviates from the power-law trend extrapolated from high-frequencies, which can be attributed to either free-free absorption (FFA) or self-synchrotron absorption (SSA) at low frequencies. An evolution is clearly revealed by comparing the set of spectra obtained on April 14, 26, and June 26, 2022, with the spectrum obtained on September 11, 2023.

\begin{figure*}
\includegraphics[width=0.48\textwidth]{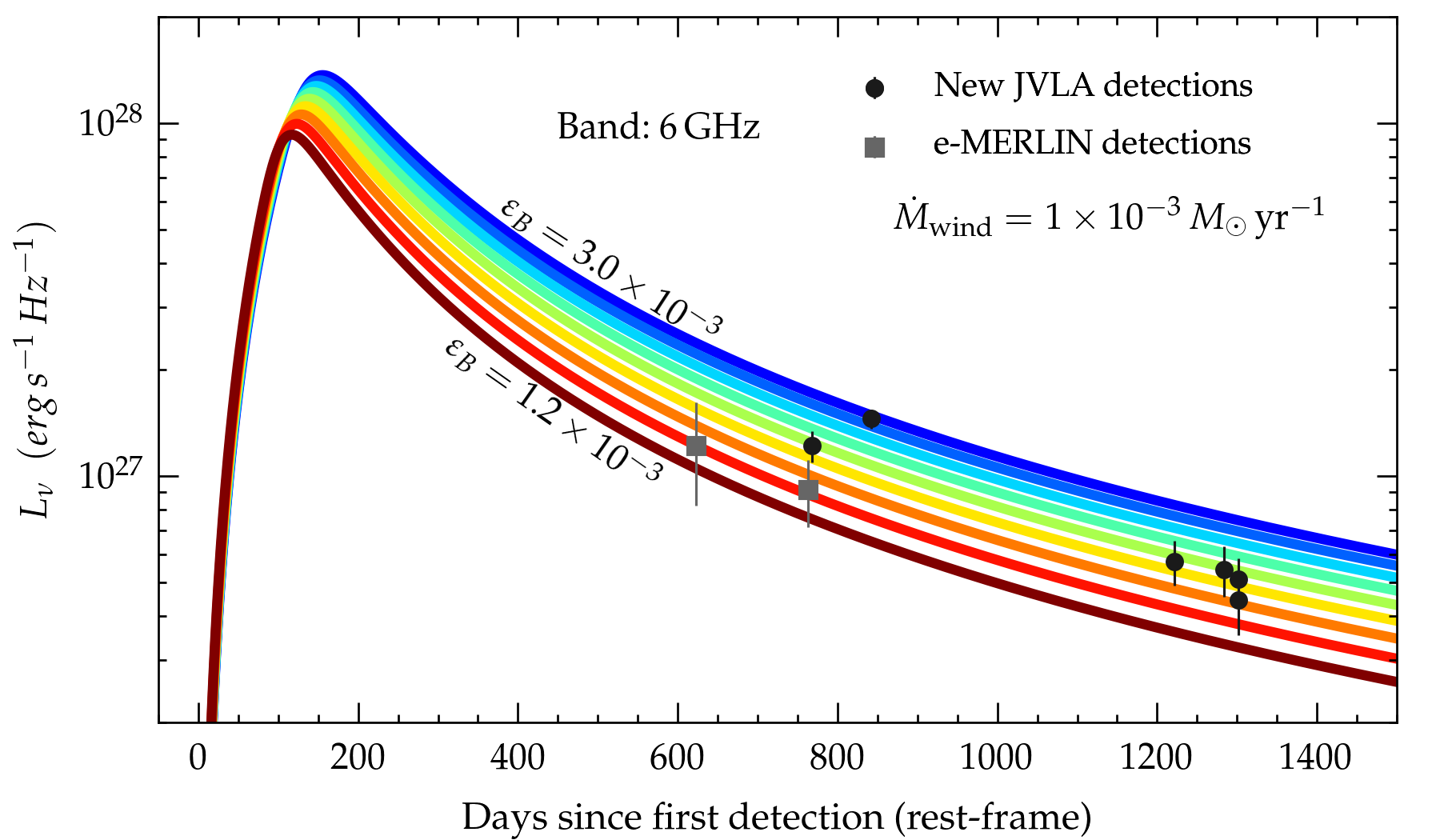}
\hfill
\includegraphics[width=0.48\textwidth]{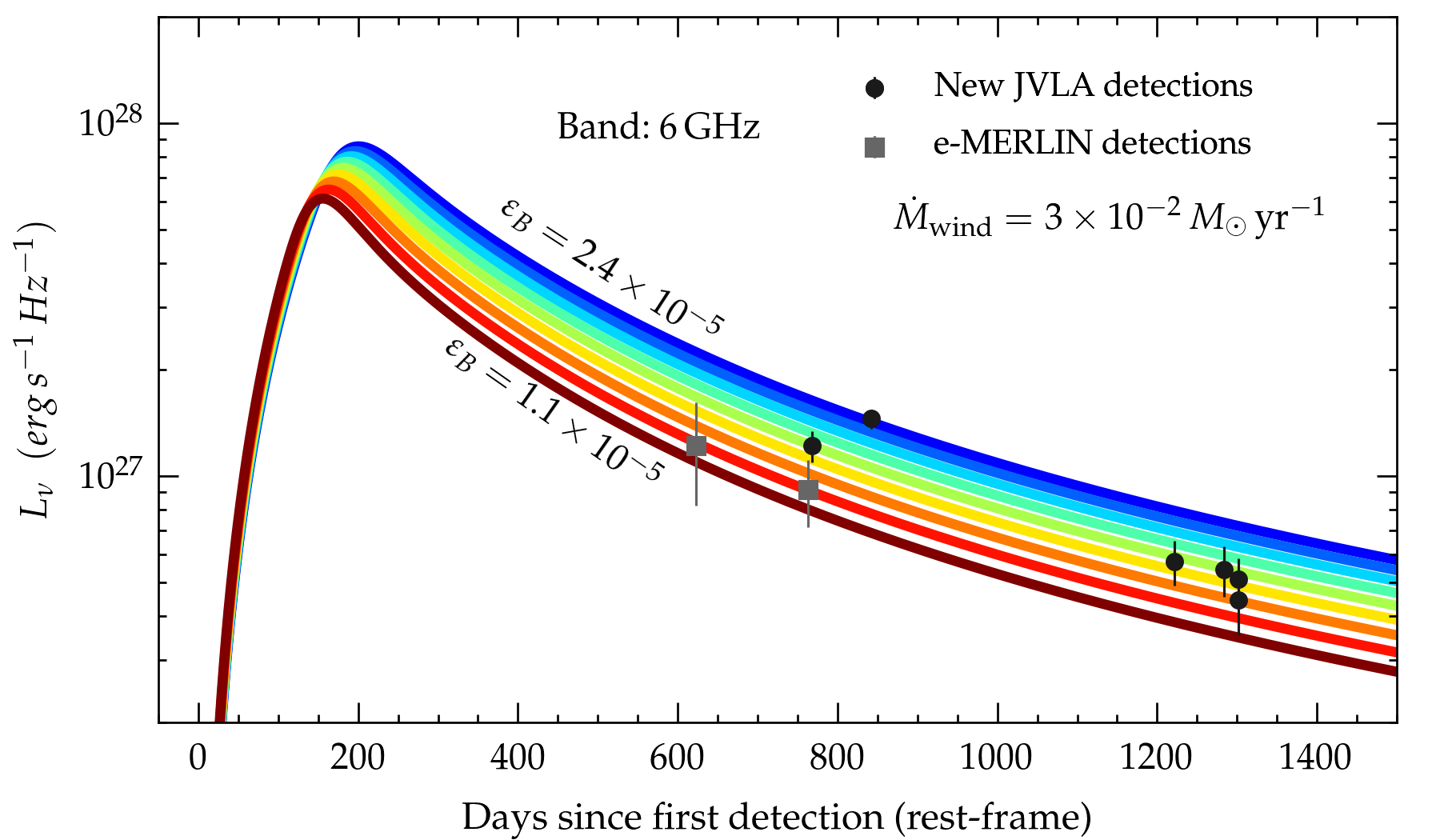}\\[0.5cm]
\vspace{0.7cm}
\includegraphics[width=0.48\textwidth]{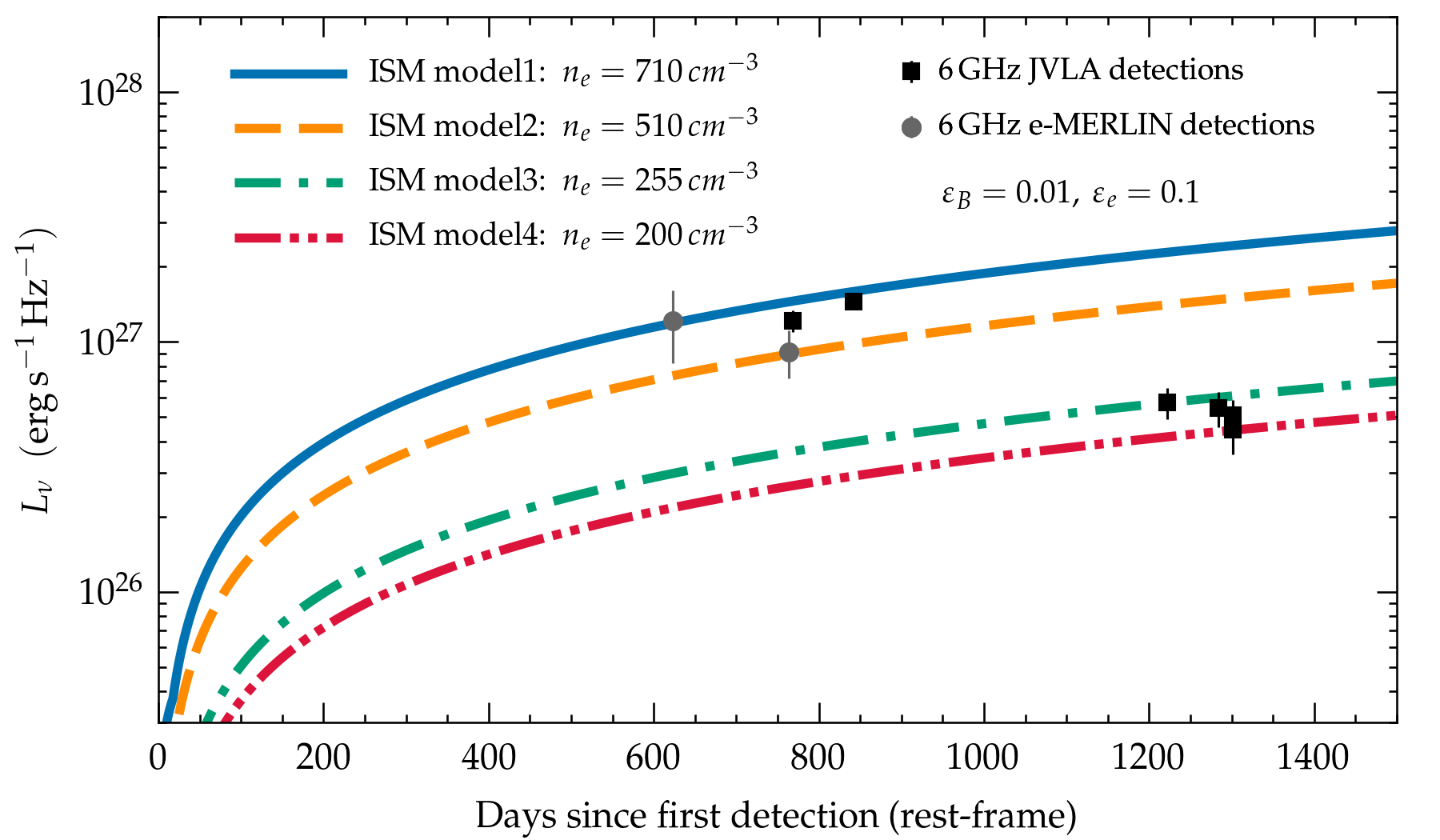}
\hfill
\includegraphics[width=0.48\textwidth]{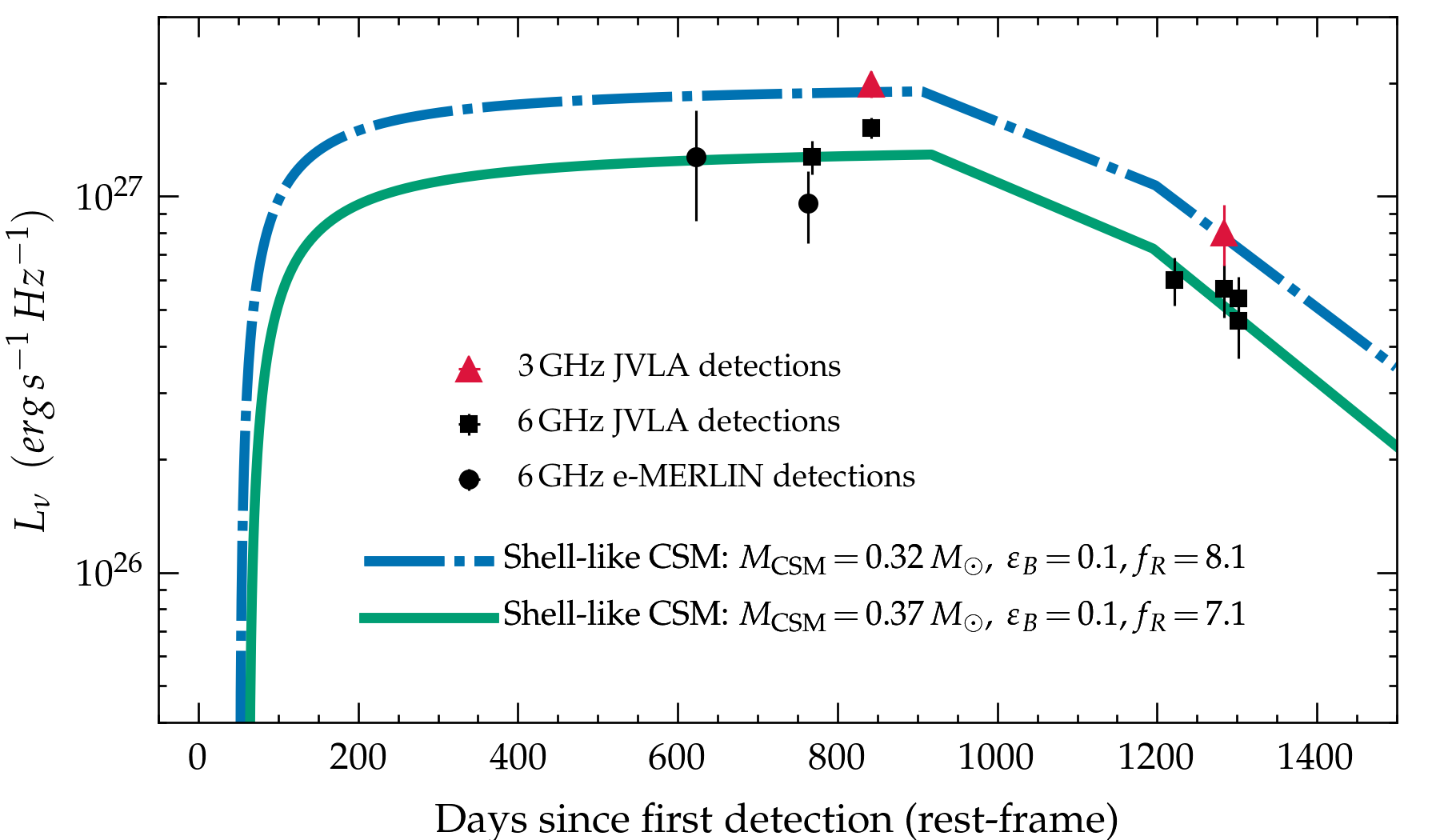}
\caption{\textbf{\textsl{Top-Left Panel:}} A wind-like CSM model constrained from the radio emissions of SN~2020eyj at 6\,GHz. We assume a mass-loss rate of $\dot{M}_{\rm wind}=1.0\times10^{-3}\,\mathrm{M_{\odot}\,yr^{-1}}$ obtained from a fit to bolometric light-curve of SN~2020eyj \citep{Kool2023} and a wind velocity of $1000\,\mathrm{km\,s^{-1}}$. We slightly adjust microphysics parameter ($\epsilon_{\rm B}$, which is indicated with different color lines) to cover all observed data points, which gives a range of $\epsilon_{\rm B}=1.2-3.0\times10^{-3}$ (from red to blue).  \textbf{\textsl{Top-Right Panel:}} similar to \textsl{Top-Left Panel}, but for a mass-loss rate of $\dot{M}_{\rm wind}=3.0\times10^{-2}\,\mathrm{M_{\odot}\,yr^{-1}}$. \textbf{\textsl{Bottom-Left Panel:}} Radio emissions of SN~2020eyj at different epochs are fitted with an analytical ISM-interaction model of \citet{Chomiuk2016}. For given standard microphysics parameters ($\epsilon_{\rm B}=0.01$ and $\epsilon_{\rm e}=0.1$), we vary the ISM densities ($n_{\rm e}$) from ${\sim}\,200\,\mathrm{cm^{-3}}$ (double dotted line) to ${\sim}\,710\,\mathrm{cm^{-3}}$ (solid line).
\textbf{\textsl{Bottom-Right Panel:}} Radio emissions of SN~2020eyj at frequencies of $3\,\mathrm{GHz}$ and $6\,\mathrm{GHz}$ fitted with a CSM-shell model given by \citet{2021ApJ...912...23H}. The best-fit models for radio detections of SN~2020eyj at $3\,\mathrm{GHz}$ and $6\,\mathrm{GHz}$ give the CSM masses of $M_{\rm CSM}\approx0.32\,M_{\odot}$ and $M_{\rm CSM}\approx0.37\,M_{\odot}$, and an interaction end time of $t_{\rm end}=909\, \mathrm{days}$ and $t_{\rm end}=917\,\mathrm{days}$, respectively. Here, $f_R = \Delta R / R_\mathrm{in}$ denotes the ratio between the thickness and the inner edge radius of the shell. \label{fig:fitting3}}
\end{figure*}

\section{Constraints on the Circumstellar Environment}\label{sec:csm}

\subsection{Wind-like CSM}

 A wind-like and hydrogen-/helium-rich CSM density structure ($\propto r^{-2}$) can be speculated from an optically thick wind (OTW) model \citep{Hachisu1999} in an SD progenitor system consisting of a WD and a hydrogen-/helium-rich donor, where the OTW is produced as the mass transfer rate exceeds the critical accretion rate (see Figure~\ref{fig:channel}). In addition to the OTW model, a SD progenitor system with a hydrogen-/helium-rich donor may drive a common envelope (CE) wind with a lower velocity of ${\lesssim}\,100\,\mathrm{km\,s^{-1}}$ than an optical thick wind (${\sim}1000\,\mathrm{km\,s^{-1}}$) when the mass transfer rate exceeds the critical accretion rate, which is called the `CE wind (CEW) model' (\citealt{Meng2017}; see also Fig.~\ref{fig:channel}). This lower wind velocity is also consistent with the typical velocities (${\sim}\,50{-}200\,\mathrm{km\,s^{-1}}$) of pre-explosion outflowing material from progenitor systems of SNe~Ia-CSM (such as SN~2006X and PTF~11kx) estimated by the detected time-varying blueshifted narrow lines \citep{Patat2007, Simon2009, Sternberg2011, Dilday2012, Maguire2013}.

In \citet{Kool2023}, they found that reproducing the late-time bolometric light curve of SN~2020eyj requires a CSM mass of $M_{\rm CSM}{\rm(wind)}\approx0.03{-}1.0\,M_{\odot}$ within a distance of $10^{17}\,\mathrm{cm}$, with the lower and upper limits corresponding to the assumed extinctions of $E(B-V)=0$ and $E(B-V)=0.5$, respectively. Assuming a steady wind with a velocity of $1000\,\mathrm{km\,s^{-1}}$, as typically adopted in the OTW model, this corresponds to a mass-loss rate of $\dot{M}_{\rm wind}\approx0.1{-}3.0\times 10^{-2}\,M_{\odot}\,\mathrm{yr^{-1}}$. For the substantially lower wind velocity of about $100\,\mathrm{km\,s^{-1}}$ expected in the CEW model \citep{Meng2017}, the inferred mass-loss rate decreases by an order of magnitude to $\dot{M}_{\rm wind}\sim0.1{-}3.0\times 10^{-3}\,M_{\odot}\,\mathrm{yr^{-1}}$. 

Following \citet{Kool2023}, we also use the wind-like CSM model and adopt the wind-loss parameters (i.e., $\dot{M}_{\rm wind}\approx0.1{-}3.0\times 10^{-2}\,M_{\odot}\,\mathrm{yr^{-1}}$ with a wind velocity of $1000\,\mathrm{km\,s^{-1}}$) given by their bolometric light-curve fitting to match the radio emissions of SN~2020eyj at 3\,GHz and 6\,GHz by adjusting the microphysics parameter (i.e., $\epsilon_{\rm B}$). The fitting leads to a range of the microphysics parameter from $\epsilon_{\rm B}\approx1.1\times10^{-5}$ to $\epsilon_{\rm B}\approx3.0\times10^{-3}$ (see top panels of Fig.~\ref{fig:fitting3}). Note that although we obtain radio observations of SN~2020eyj at frequencies of $1.5\,\mathrm{GHz}$ and $10\,\mathrm{GHz}$ as well, we do not provide a fit to these two frequencies because each frequency has one JVLA detection only and $1.5\,\mathrm{GHz}$ emission is in the optically thick regime. 

Given the uncertainties of wind velocity and the microphysics parameter, we further use the radio observations of SN~2020eyj at the 6\,GHz as the representative case to explore the allowed parameter space of the mass-loss rate and wind velocity following the method of \citet{Chomiuk2012}. We find that reproducing the observed $6\,\rm{GHz}$ radio evolution requires a wind parameter of $A=\dot{M}_{\rm wind}/\upsilon_{\rm wind}$, expressed in units of  $10^{-6}\,M_{\odot}\,\mathrm{yr^{-1}}/100\,\mathrm{km\,s^{-1}}$, with values of $5.8$, $29.2$, $150.0$ and $768.3$ for a range of microphysics parameters of $\epsilon_{\rm B}=0.1$, $0.01$, $0.001$ and $0.0001$, respectively.

\begin{figure}
\centering
\includegraphics[width=0.47\textwidth]{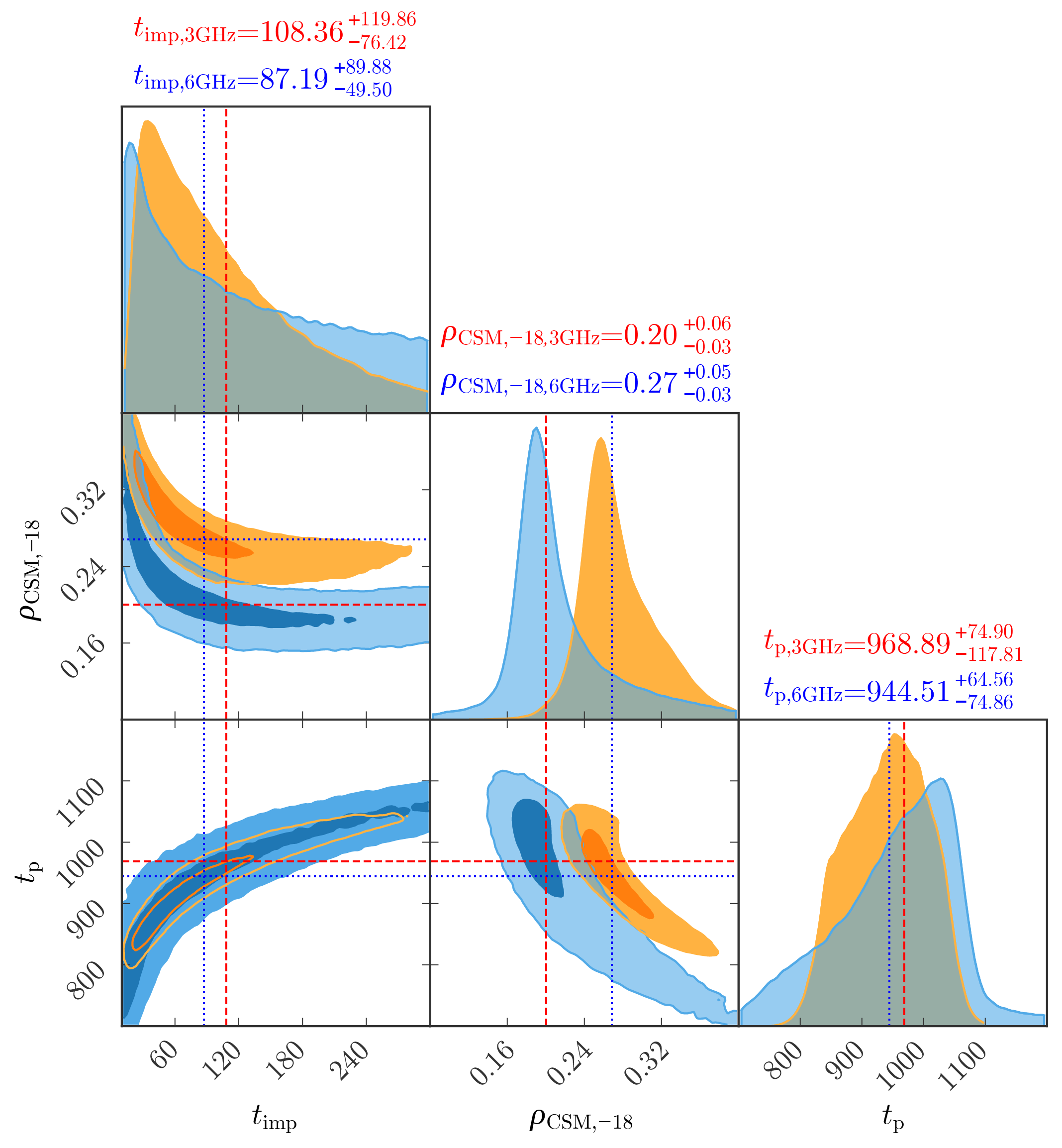}
\caption{Posterior probability distribution results of MCMC fitting of radio light curves with a CSM-shell model. The blue is for 3\,GHz and the yellow is for 6\,GHz. The 50 percentiles are marked in each of the counters and histograms as the best-fitting values, where the red dashed lines are for 3\,GHz, and the blue dashed lines are for 6\,GHz. We take 16\% and 84\% of the distributions as the lower and upper limits, respectively.}
\label{fig:fitlc}
\end{figure}

\subsection{Shell-like CSM}
Alternatively, a constant density CSM-shell configuration is expected to form from pre-explosion mass ejection, such as nova events \citep{Dilday2012, Chomiuk2012} or the ejection of a common envelope layer (\citealt{Meng2017, Soker2023}; see Fig.~\ref{fig:channel}). Observations of the SN Ia event PTF~11kx indicate the presence of multiple CSM shells interacting with the SN ejecta \citep{2021ApJ...912...23H}. Some observational evidence also suggests that a portion of SNe~Ia-CSM may have a similar CSM-shell configuration as in PTF~11kx \citep{Chomiuk2012}.  In \citet{2021ApJ...912...23H}, they have explored the radio emission of SN Ia ejecta interacting with circumstellar shells (shell-like CSM), which differs significantly from the wind-CSM model. We adopt their model with $\epsilon_e=0.1$ and $\epsilon_B=0.1$ to fit the radio emissions of SN~2020eyj detected at the two frequencies of $3\,\mathrm{GHz}$ and $6\,\mathrm{GHz}$ (see Fig.~\ref{fig:fitting3}). We also modify the original hydrogen-dominated CSM composition to a helium-rich medium, assuming fully ionized helium with  $n_e=\rho/(2m_p)$. We employ a Markov Chain Monte Carlo (MCMC) method (see Fig.~\ref{fig:fitlc}) to explore the model parameter space. No additional constraints are imposed except for on the interaction end time, $t_{\rm end}$, which is restricted to $623-1300$ days. The lower bound corresponds to the first radio detection, when the interaction was still ongoing. The upper bound is motivated by the late-time decline in radio emission around $1300$ day. As shown in Fig.~\ref{fig:fitlc}, the resulting posterior distributions are well defined, suggesting that the model parameters are reasonably constrained.

\begin{deluxetable}{lcrr}
\tablewidth{0pt}
\tablecaption{Radio light curve fitting parameters of \obj\ with a CSM-shell model. \label{tab:lcfit}}
\tablehead{\colhead{Parameter} & \colhead{Unit} & \colhead{3~GHz} & \colhead{6~GHz}}
\colnumbers
\startdata
$t_\mathrm{imp}$                &  (Day)                                          &     $54^{+59}_{-37}$           &    $63^{+49}_{-32}$          \\
$\rho_\mathrm{CSM}$      &  ($\times10^{-18}\,\mathrm{g}\,\mathrm{cm}^{-3}$)     &     $0.12^{+0.05}_{-0.02}$     &    $0.15^{+0.03}_{-0.02}$    \\
$t_\mathrm{end}$                    &  (Day)                                          &     $909^{+77}_{-125}$         &    $917^{+51}_{-63}$         \\
$R_\mathrm{in}$          &  ($\times10^{16}\,\mathrm{cm}$)                       &     $1.18^{+0.73}_{-0.65}$     &    $1.28^{+0.61}_{-0.50}$    \\
$\upsilon_\mathrm{ej}$          &  ($\mathrm{km}\,\mathrm{s}^{-1}$)               &     $25318^{+11497}_{-5818}$   &    $23565^{+5985}_{-4056}$   \\
$f_\mathrm{R}$                    &                                                 &     $8.1^{+11.4}_{-3.7}$       &    $7.1^{+5.3}_{-2.7}$       \\
$M_\mathrm{CSM}$         &  ($\mathrm{M_\odot}$)                           &     $0.32^{+0.10}_{-0.05}$     &    $0.37^{+0.03}_{-0.02}$    \\
\hline
\enddata
\end{deluxetable}

Interestingly, a shell-like CSM configuration can reproduce the detailed radio light curves at both $6\,\mathrm{GHz}$ and $3\,\mathrm{GHz}$ with a consistent set of model parameters (see the bottom-right panel of Fig.~\ref{fig:fitting3}). In particular, the $6\,\mathrm{GHz}$ radio light curve is best reproduced with a magnetic-field energy fraction of $\epsilon_{\rm B}=0.1$ and a shell-width parameter of $f_R=\Delta R/R_{\rm in}\approx7.1$, where $\Delta R$ and $R_{\rm in}$ denote the shell width and inner radius, respectively. The corresponding best-fit CSM-shell mass is $M_{\rm CSM,shell}\approx0.37\,M_{\odot}$ (see Fig.~\ref{fig:fitting3}). Similarly, fitting the $3\,\mathrm{GHz}$ radio light curve yields a best-fit CSM-shell mass of $M_{\rm CSM,shell}\approx0.32\,M_{\odot}$, which is consistent with the value inferred from the $6\,\mathrm{GHz}$ observations. Furthermore, fitting the radio light curves at $3\,\mathrm{GHz}$ and $6\,\mathrm{GHz}$ with the shell-like CSM model yield explosion shock velocities of $25318^{+11497}_{-5818}$~km\,s$^{-1}$ and $23565^{+5985}_{-4056}$~km\,s$^{-1}$, respectively. Assuming a plausible efficiency for converting the kinetic energy of the SN ejecta into radiation through CSM interaction, the best-fit shell masses inferred from the radio observations can also reproduce the late-time bolometric light curve of SN~2020eyj between days $+50$ and $+250$ after discovery (see Fig.~\ref{fig:shell_lc} of Section~\ref{app:lc}). These results suggest that the progenitor system of SN~2020eyj was surrounded by a substantial CSM shell prior to explosion (see Fig.~\ref{fig:channel}).

\subsection{ISM-like CSM}
An interstellar medium (ISM)-like environment (extended region of constant density) may arise in both SD and DD progenitor scenarios. In the classical DD scenario, an SN~Ia explosion is triggered by the merger of two WDs \citep{Webbink1984, Iben1984}. In the absence of substantial pre-explosion mass ejection, the progenitor system is expected to be surrounded by a low-density environment similar to the ISM. The ``spin-up/spin-down model'' within the SD framework \citep{DiStefano2011,Justham2011,Hachisu2012} provides another possible pathway to an ISM-like environment. In this scenario, the WD accretes mass and angular momentum from its companion, leading to rapid rotation that increases the critical mass for explosion. The explosion may therefore be delayed until the WD loses sufficient angular momentum through spin-down. The spin-down timescale can be as long as $10^{6}-10^{9}\,\rm{yrs}$ \citep{Ilkov2012}. Such a long spin-down timescale allows previously ejected CSM to disperse, leaving the progenitor system surrounded by an ISM-like environment at the time of explosion.

However, reproducing the observed radio emission of SN~2020eyj with an ISM-interaction model requires a density substantially higher than that of the typical ISM (see Fig.~~\ref{fig:fitting3}). Moreover, the predicted radio luminosity in an ISM-like environment generally increases with time \citep[e.g.,][]{Chomiuk2012, Perez-Torres2014, Kundu2017, Lundqvist2020}, which is in contrast to the overall declining trend observed in the radio light curves of SN~2020eyj. Furthermore, the presence of strong helium lines in its optical spectra suggest a helium-rich circumstellar environment rather than a typical ISM environment \citep{Osterbrock2006}. Therefore, these findings strongly disfavor an ISM-like environment as the origin of the observed radio emission from SN~2020eyj.

\begin{figure}
\centering
\includegraphics[width=0.49\textwidth]{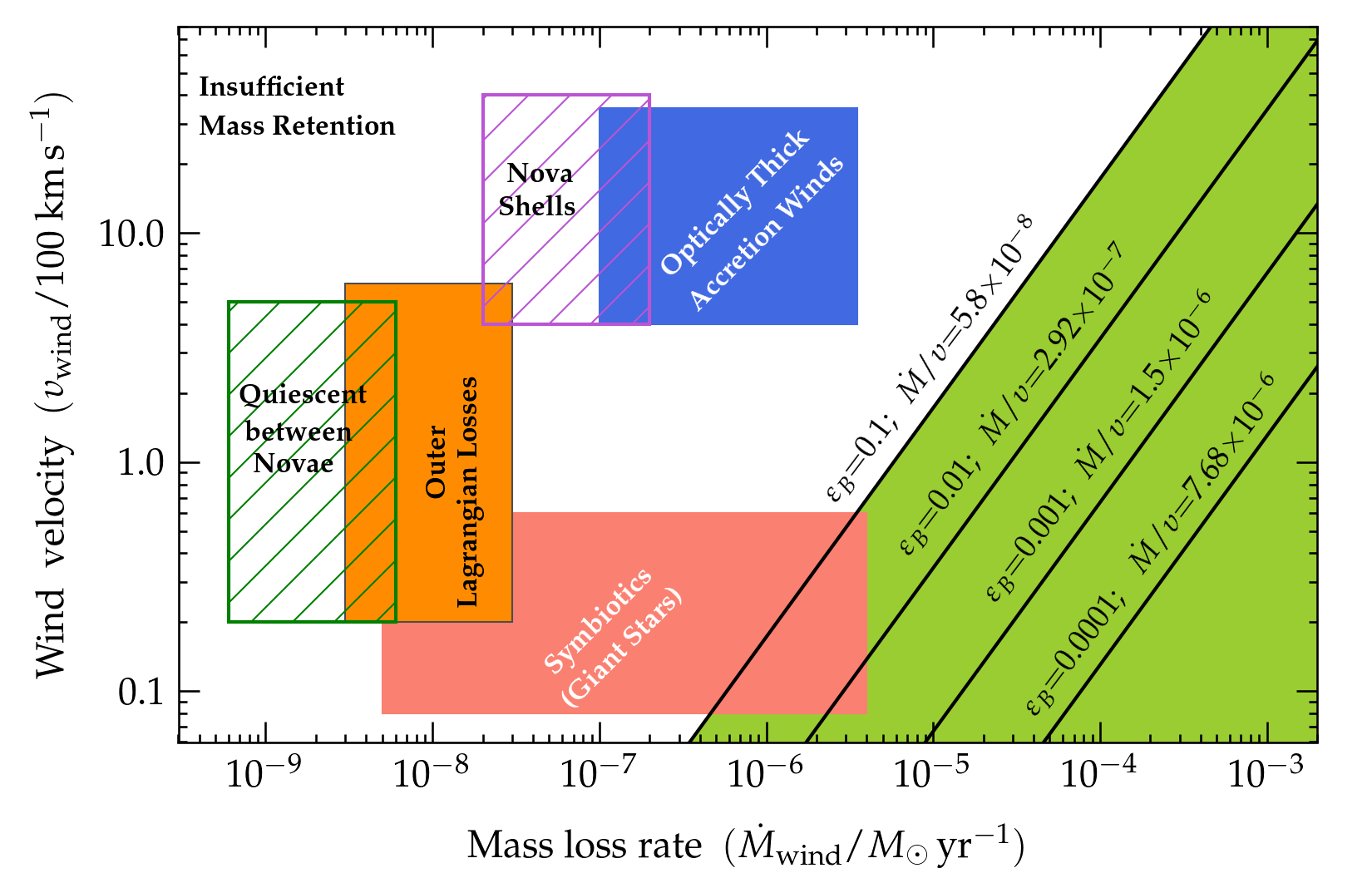}
\caption{Constraints on wind-like CSM models in the mass-loss rate and wind velocity space for different progenitor channels by adopting the same method as \citet{Chomiuk2012}. The oblique black solid lines represent wind mass-loss rates derived from the constraints imposed by the radio evolution of SN~2020eyj at $6\,\mathrm{GHz}$ for the cases of $\epsilon_{\rm B}=0.1$, $0.01$, $0.001$ and $0.0001$. The lower-right regions, shaded in green, denote the allowed parameter space for explaining radio observations of SN~2020eyj.} \label{fig:progenitors}
\end{figure}

\section{Implications for the Progenitor System}\label{sec:discussion}

In the DD scenario, several mechanisms have recently been proposed to explain the ejection of material from the progenitor system prior to the SN Ia explosion, potentially giving rise to the surrounding CSM. These include $\textit{tidal tail ejection}$ \citep{Raskin2013}, $\textit{mass outflows during rapid accretion}$ \citep{Guillochon2010, Dan2011}, $\textit{disk winds}$ \citep{Ji2013} and $\textit{mass ejection prior to violent mergers}$ \citep{Pakmor2013, Shen2013}. However, these mechanisms generally predict relatively small amounts of CSM, making it difficult to produce the substantial CSM mass (${\sim}\,0.03{-}1.0\,M_{\odot}$) required to simultaneously explain the observed optical light curve and long-term radio emission of SN~2020eyj. Consequently, these DD progenitor models appear to be disfavored for SN~2020eyj.

For a wind-like CSM, a good fit to the bolometric light-curve tail and long-term radio evolution of SN~2020eyj requires a pre-explosion mass-loss rate of $\dot{M}_{\rm wind}\approx0.1{-}3.0\times 10^{-2}\,M_{\odot}\,\mathrm{yr^{-1}}$, assuming a wind velocity of $1000\,\mathrm{km\,s^{-1}}$, as adopted in OTW model. These mass-loss rates exceed those predicted by the OTW model approximately $1$--$2$ orders of magnitude (\citealt{Wang2009, Moriya2019}; see Fig.~\ref{fig:channel}), leaving only a limited region of the progenitor parameter space viable as shown in Fig.~\ref{fig:progenitors}. For the CEW model, which predicts a substantially lower wind velocity of $\upsilon_{\rm wind}=100\,\mathrm{km\,s^{-1}}$, the required mass-loss rates decrease by an order of magnitude to  $0.1{-}3.0\times10^{-3}\,\mathrm{M_{\odot}\,yr^{-1}}$. However, such high mass-loss rates remain challenging to reconcile with the predictions of the CEW model \citep{Meng2017, Cui2022}. In addition, we find that while wind-like CSM (CEW) models can broadly reproduce an overall declining radio light curve, but they significantly deviate from the observed radio light curves of SN~2020eyj at both 3 and 6\,GHz (see Fig.~\ref{fig:fitting3}), particularly in the evolution slope and the bump observed around day +841 at 6\,GHz. This may suggest that a simple steady-wind density profile is hard to adequately describe the circumstellar environment of SN~2020eyj.

Multiple CSM-shells have been confirmed in the SN~Ia-CSM event PTF~11kx by observing time-varying narrow absorption lines in high-resolution spectra \citep{Dilday2012}. CSM-shells in PTF~11kx were originally explained by a symbiotic nova progenitor in a SD model similar to RS~Ophiuchi \citep{Dilday2012}, in which the multiple-shell CSM structures originated from recent recurrent nova eruptions whose ejecta sweep the wind of a giant companion star. Aside from the presence of helium emission lines in the late-time spectra of SN~2020eyj, whereas PTF~11kx shows narrow hydrogen lines, SN~2020eyj shares many similarities with PTF~11kx in the early optical light curves (${\sim}\,50\,\mathrm{days}$), spectral features, and their evolution. On the other hand, helium novae have long been theoretically predicted \citep{Kato1989}, with one known candidate, V445 Puppis \citep{Kato2003, Kato2008, Nyamai2021,  Schaefer2024}. Such systems have also been proposed as potential progenitors of SNe~Ia through helium accretion onto WDs \citep{Kato2003, Kato2008, Kemp2021}. Therefore, a helium-nova progenitor scenario may provide a natural explanation for the helium-rich composition and shell-like structure of the CSM surrounding SN~2020eyj. 
However, typical helium-nova eruptions are unlikely to eject sufficient material to produce the massive, dense CSM shell \citep{Dilday2012} required to reproduce the observed light curves of SN~2020eyj (see Fig.~\ref{fig:progenitors}).

The ejection of a CE layer prior to explosion is expected to produce a shell-like CSM \citep{Meng2017, Soker2023}. In the SD scenario with a helium-star donor, the donor fills its Roche lobe in the Hertzsprung gap or on the first giant branch and transfers helium-rich material to the WD on a thermal timescale. When the mass-transfer rate exceeds the critical accretion rate, a helium-rich CE forms around the binary instead of an optically thick wind, and part of this envelope is lost as a slow wind with ${\lesssim}\,100\,\mathrm{km\,s^{-1}}$ \citep{Meng2017}. By the time the WD approaches the Chandrasekhar limit, a helium-rich CE of ${\sim}\,0.3{-}0.75\,M_{\odot}$ remains \citep{Meng2026}, depending on the initial WD and helium-star masses, and this envelope is expected to be dynamically ejected with a velocity of order $100\,\mathrm{km\,s^{-1}}$ before the explosion \citep{Cui2022, Meng2026}. The predicted mass of helium CE ejected after the CEW phase is comparable to the CSM shell mass of $M_{\rm CSM}\approx0.37\,M_{\odot}$ inferred from the radio light curves and to the $0.03{-}1.0\,M_{\odot}$ required by the bolometric light curve \citep{Kool2023}. However, the delay between the CE ejection and the explosion is poorly constrained in CEW channel, and the physical mechanism that ejects the CE is still unknown. For $R_{\rm in}\approx1.3\times10^{16}$\,cm and $\Delta R\approx9\times10^{16}$\,cm, ejecta at $100\,\mathrm{km\,s^{-1}}$ require a delay of $\gtrsim40$\,yr to reach the inner edge. They also require ejection spread over $\gtrsim290$\,yr, or a velocity spread up to ${\sim}\,800\,\mathrm{km\,s^{-1}}$, to produce a shell of this width (see Figure~\ref{fig:channel}).

The recently developed core-degenerate (CD) scenario in a DD system \citep{Soker2023} also predict a CE layer in the progenitor system. In this scenario, it involves two CE phases, during which the hydrogen-rich envelope of the AGB is removed during the first CE evolution (CEE). After a time delay, the WD enters the second CEE with a helium-rich layer as the helium-rich core expands, and eventually merges with the carbon-oxygen core to trigger a SN Ia, ejecting the helium-rich envelope into the surrounding environment with a mass ${\sim}\,0.03{-}1.0\,M_{\odot}$ and a velocity of ${\lesssim}\,100\,\mathrm{km\,s^{-1}}$ \citep{Soker2023}. Differing from the original CD model involving the merger of a WD with the core of an AGB companion star during a CE evolution (CEE), which predicts a hydrogen-rich CSM environment \citep{Ilkov2012, Ilkov2013, Soker2013, Aznar-Siguan2015}, the new CD model forms a helium-rich shell-like CSM (see Figure~\ref{fig:channel}). However, for $R_{\rm in}\approx1.3\times10^{16}$\,cm and $R_{\rm out}\approx1.0\times10^{17}$\,cm, CE ejecta with $v\lesssim100\,\mathrm{km\,s^{-1}}$ require look-back times of $\gtrsim40$\,yr for the inner edge and $\gtrsim330$\,yr for the outer edge. This is in tension with the merger-to-explosion delay of only weeks to tens of years allowed in the CD scenario \citep{Soker2023}, unless the ejecta span a wide range of velocities.

It has been speculated that a massive helium-rich CSM in a long-lived disk could be built up around the binary system prior to nova eruptions. This scheme overcomes the mass deficit challenge in the CSM-shell model involving nova eruptions and provides an alternative explanation for SN~2020eyj in the context of the SD scenario \citep{Kool2023}. Evidence for such a long-lived disk comes from radio observations of the only known helium nova, V445~Puppis \citep{Kato2008, Nyamai2021}. The CSM disk model can also explain the luminous infrared counterpart of SN~2020eyj \citep{Kool2023}. However, whether such helium novae can ultimately lead to an SNe Ia remains uncertain (\citealt{Schaefer2024}, but see \citealt[][]{Kato2003, Kato2008, Kemp2021}). Furthermore, the exact mass of such long-lived disks remains poorly constrained from both theoretical and observational perspectives \citep{Soker2023}. To strengthen the case for massive long-lived disks, additional observations are needed to identify V445~Puppis-like events.

\section{Conclusions and Summary}\label{sec:summary}

We have presented JVLA monitoring of \obj\ at 1.5--10\,GHz between days $+768$ and $+1301$ after discovery, including the first multi-epoch radio spectra of a SN~Ia. The 6\,GHz light curve declines overall after a steady phase between days 623 and 841, and the spectra show a low-frequency turnover consistent with absorbed synchrotron emission. We explored proposed formation channels for SNe Ia and found that the existing models seem challenging to explain the constrained CSM properties of SN~2020eyj (see Fig.~\ref{fig:channel}). Our main results and conclusions can be summarized as follows: 
\vspace{-\topsep}

\begin{itemize}\itemsep5pt 
\item[(1)] The observed long-term radio evolution of SN~2020eyj disfavors an ISM-like circumstellar environment. Moreover, traditional double-degenerate (DD) progenitor scenarios are unlikely to produce sufficient CSM mass to account for the observed radio emission.

\item[(2)] A spherically symmetric wind-like CSM model reproduces the overall decline of the radio light curves but fails to capture their detailed temporal evolution, particularly the observed decline slope and the rebrightening around day $+841$ at $6\,\rm{GHz}$. Moreover, the exceptionally high pre-explosion mass-loss rates required to explain the radio observations of SN~2020eyj pose significant challenges to existing SN~Ia progenitor scenarios.

\item[(3)]A helium-rich, constant-density CSM shell of with a mass of about $0.3$--$0.4\,M_{\odot}$ can simultaneously reproduce the observed radio and bolometric light curves and account for the helium-rich features in the optical spectra. These findings provide strong evidence for substantial mass loss of helium-rich material from the progenitor system prior to explosion.

\item[(4)] The inferred CSM properties suggest that the ejection of a helium-rich CE layer prior to explosion may provide the most plausible explanation among the progenitor scenarios considered in this study, although significant theoretical uncertainties remain. In particular, further investigations are needed to understand the physical mechanisms responsible for CE ejection shortly before explosion in both the CEW and CD scenarios. Also, the effects of CSM asymmetries on the radio emission arising from the CSM-interaction are investigated in a separate paper (R. Chiba et al., in preparation).

\end{itemize} 

Further development of new progenitor and explosion models may still be needed to explain the radio observations of SN~2020eyj presented in this work. In particular, it remains an important open question whether helium-nova systems can produce a disk-like CSM configuration \citep{Kato2008, Nyamai2021} with sufficient mass and density to provide an alternative channel to account for the observed radio emissions here. Furthermore, SN~2020eyj remains to be the only SN~Ia for which the long-term radio evolution has been characterized, raising the possibility that it was originated from an unusual progenitor system or represents a distinct subclass of SNe~Ia. However, a range of radio luminosities remains unexplored for SNe~Ia owing to the limited sensitivity of current radio facilities. Future radio observations with next-generation facilities, such as the Square Kilometre Array (SKA) and the next-generation Very Large Array (ngVLA), will enable more sensitive searches for radio emission from SNe~Ia, providing crucial constraints on their circumstellar environments and potential progenitor channels.

\begin{acknowledgments}
We thank Takashi Moriya, Lifan Wang, Luqian Wang, and many others for very fruitful discussions on this project. This work is supported by the CAS Project for Young Scientists in Basic Research (YSBR-148), the National Natural Science Foundation of China (NSFC, Nos.\ 12288102, 12125303, 11873016, 12041301, 12121003, 12333008, 12303047), the Yunnan Revitalization Talent Support Program “YunLing Scholar” project, the Strategic Priority Research Program of the Chinese Academy of Sciences (grant Nos. XDB1160303, XDB1160300), the National Key R\&D Program of China (Nos.\ 2021YFA1600401, 2021YFA1600400, and 2021YFA0718500), the Chinese Academy of Sciences (CAS), International Centre of Supernovae (ICESUN), Yunnan Key Laboratory of Supernova Research (No.~202505AV340004), Yunnan Fundamental Research Projects (202401BC070007), the New Cornerstone Science Foundation through the XPLORER PRIZE and the ``Yunnan Revitalization Talent Support Program'' -- Science \& Technology Champion Project (No. 202305AB350003). The work of FR is supported by the Klaus Tschira Foundation, by the Deutsche Forschungs-
gemeinschaft (DFG, German Research Foundation) – RO 3676/7-1, project num-
ber 537700965, and by the
European Union (ERC, ExCEED, project number 101096243).
Views and opinions expressed are, however, those of the
authors only and do not necessarily reflect those of the
European Union or the European Research Council Executive
Agency. Neither the European Union nor the granting
authority can be held responsible for them..

This work made use of data from the Karl G.\ Jansky Very Large Array (JVLA), which is operated by the NRAO. The National Radio Astronomy Observatory (NRAO) is a facility of the U.S. National Science Foundation operated under cooperative agreement by Associated Universities, Inc. This project used data obtained with the Dark Energy Camera (DECam), which was constructed by the Dark Energy Survey (DES) collaboration.
\end{acknowledgments}

\facilities{VLA, $e$-MERLIN, PO:1.2m (ZTF), Blanco (DECam), PS1, Gaia}

\software{CASA \citep{2007ASPC..376..127M}, DIFMAP \citep{1997ASPC..125...77S}}

%\clearpage
\appendix
\twocolumngrid

\section{JVLA Observations and Data Reduction}\label{app:obs}
Six observations of \obj\ were conducted with the Karl G. Jansky Very Large Array (JVLA, twenty-seven 25-meter radio telescopes). All observations were performed using the most extended A configuration, with a maximum baseline 36 km, and employed J1125+2610 (11$^h$25$^m$53$^s$.712, 26$^\circ$10$^{\prime}$19$^{\prime\prime}$.979, approximately $4.4^\circ$ from the target) as a nearby calibrator. The initial observations were carried out on April 14, April 26, and June 26, 2022, under project codes 22A-472 and 22A-497 (PI: Javier Moldon). The subsequent four epochs were observed in monitoring mode on July 11, September 11, and September 29, 2023, at 13:47 (UT) and 21:48 (UT), under project codes 23A-425 and 23B-299 (PI: Xiaolong Yang). All observations were conducted in dual-polarization mode. The JVLA observations recorded signals across a broad frequency range, including the L-band (1–2 GHz), S-band (2–4 GHz), C-band (4–8 GHz), and X-band (8–12 GHz). Further details are provided in Table~\ref{tab:jvla}.

We manually calibrated all datasets using the Common Astronomy Software Application (CASA v6.4.1.12) \citep{2007ASPC..376..127M}. The data reduction followed standard procedures described in the CASA cookbook\footnote{\url{https://casa.nrao.edu/casadocs}}. Prior to calibration, an ionospheric correction was applied using data retrieved from the CDDIS archive\footnote{\url{https://cddis.nasa.gov/}}. The overall flux density scale for the primary flux calibrator was established based on the `Perley-Butler 2017' flux density standard \citep{2017ApJS..230....7P}. Antenna-based delay and bandpass solutions were derived and subsequently applied to the target data. Finally, complex gain solutions were determined using the nearby calibrator J1125$+$2610 and bootstrapped to the target.

\begin{figure}
\centering
\includegraphics[width=0.46\textwidth]{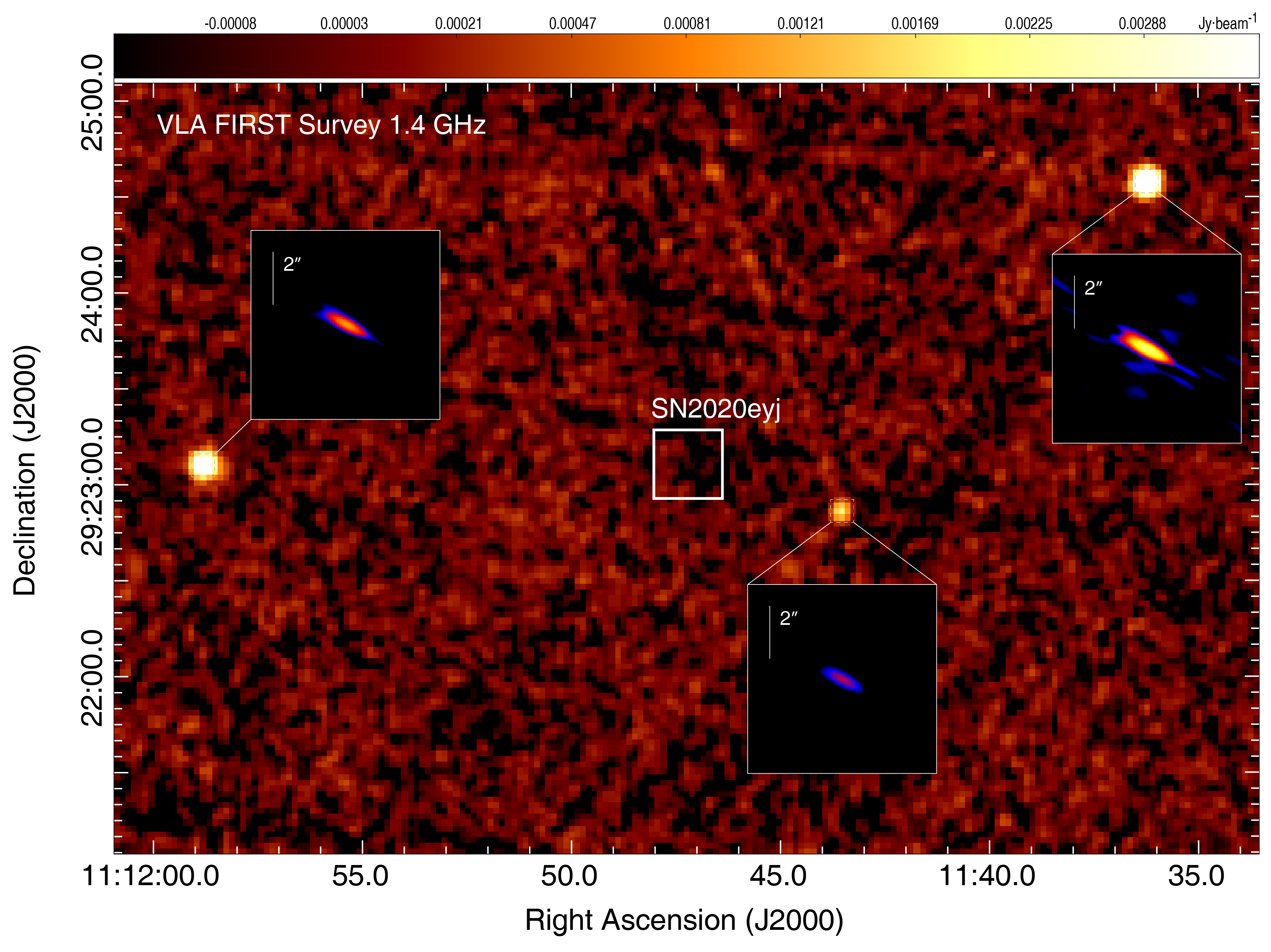}
\caption{The in-beam calibrators. The background image is the VLA FIRST survey 1.4\,GHz where the three in-beam calibrators are visible. In the sub-panel aside from each calibrator, we depict its JVLA 6\,GHz image obtained from the observation 22A-472.}
\label{fig:calibrator}
\end{figure}

The calibrated data were imaged using the \textsc{tclean} task in CASA, incorporating the efficient $w$-term and facetting (`$gridder = $`$widefield$') to account for non-coplanar array effects in wide-field imaging. Imaging was conducted with a grid size of $16384\times16384$ pixels, ensuring the deconvolution and modeling of confusing sources in the sidelobes of the primary beam. The Briggs weighting scheme ($robust = 0.0$) was employed to suppress sidelobes in the synthesized beam. Within the 10$^{\prime}$ field of view, three bright (SNR $>$ 100) in-beam radio sources—designated as calibrators A, B, and C (see Figure~\ref{fig:calibrator})—were identified, enabling self-calibration in CASA. Imaging used the MT-MFS (multi-term multi-frequency synthesis) algorithm with a two-term spectral model approximation ($nterms = 2$). Upon completing the cleaning process, the final model was written back to the MODEL\_DATA column of the measurement set (MS) for use in self-calibration. Self-calibration was initially performed on phase-only corrections, followed by adjustments to both phase and amplitude.

\begin{figure}
\centering
\includegraphics[width=0.47\textwidth]{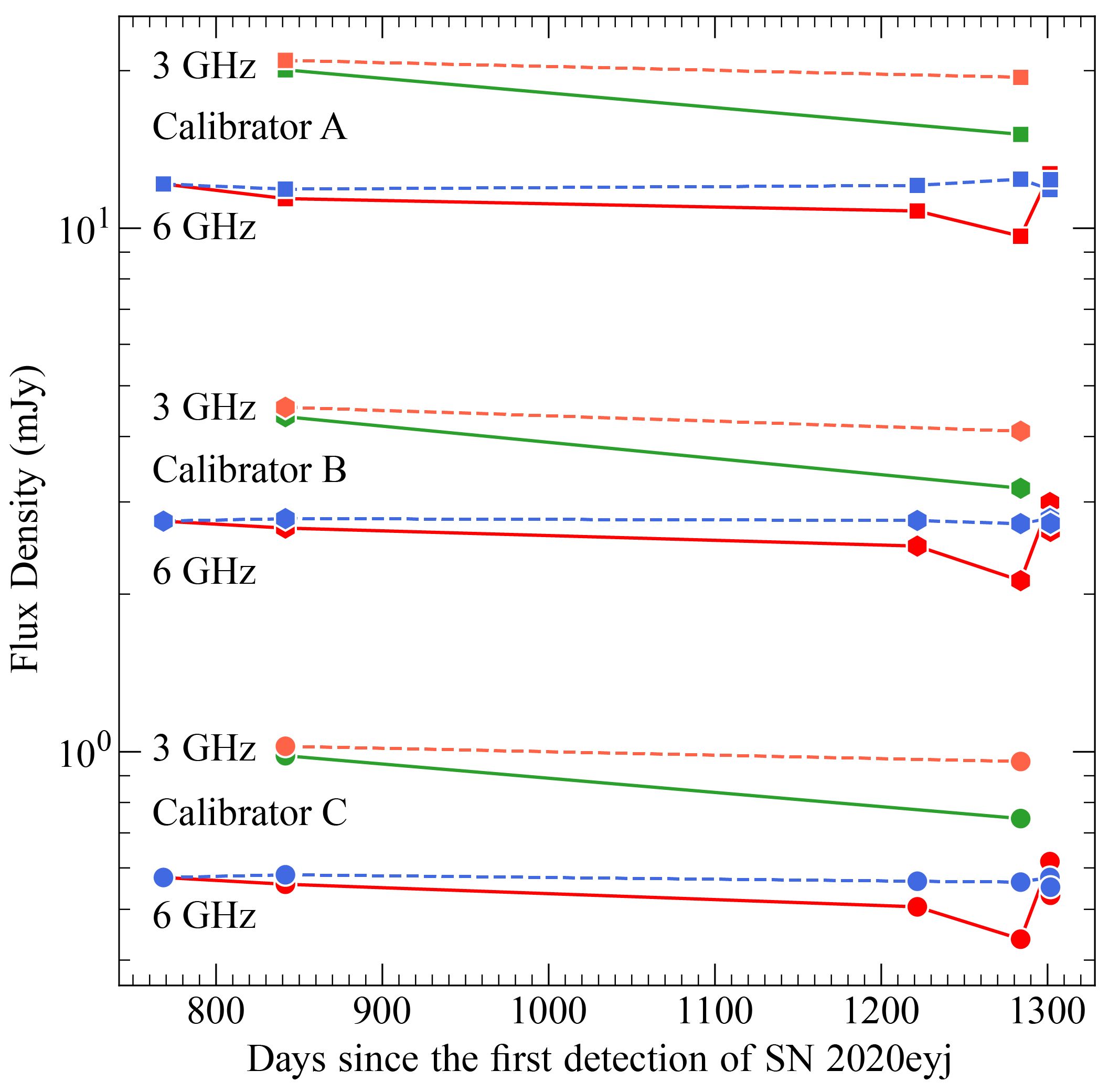}
\caption{Radio flux densities of three in-beam calibrators. The solid lines are the originally detected radio flux densities, while the dashed lines represent the radio flux densities after gain corrections.}
\label{fig:gain}
\end{figure}

The uncertainties in flux density measurements from JVLA observations are primarily attributed to thermal noise and calibration errors in amplitude, where the calibration errors result from the bootstrapping of flux densities from calibrators to the target. The absolute flux density scale for the JVLA observations was determined using calibrators with known flux densities. In this monitoring campaign, 3C~48 (observations 22A-472 and 22A-497) and 3C~286 (observations 23A-425 and 23B-299) served as the flux density scale calibrators. We employed flux density scale calibrator models provided by CASA, acknowledging that potential errors may arise from flux density variability over time, frequency, and elevation. The accuracy of the flux density scale calibration is estimated to be within 5\% across our observational bands\footnote{\url{https://science.nrao.edu/facilities/vla/docs/manuals/oss2023B}}. To mitigate calibration errors, we applied an overall scaling using three above mentioned in-beam calibrators. Notably, strong variability correlations among the three in-beam calibrators (see Figure~\ref{fig:gain}) suggest significant residual gain errors in calibration. To address this, we determined a relative gain correction by referencing the first epoch observation on April 14, 2022 (see calibrator flux densities in Table~\ref{tab:calibrator}). This correction was subsequently applied to the target to derive the final flux densities for the JVLA observations (see Table~\ref{tab:jvla}). However, this relative gain correction was not applied to the initial two $e$-MERLIN observations, as flux density information for the three calibrators was unavailable for those epochs.

The JVLA observations span a wide bandwidth, allowing the determination of the in-band radio spectral index $\alpha$ ($f_p \propto \nu^\alpha$, see Table~\ref{tab:jvla}). This was achieved using DIFMAP \citep{1997ASPC..125...77S} by fitting a point source model, except for the L-band observation on April 26, 2022, and the S-band observation on September 11, 2023. These specific observations were too faint to perform reliable model fitting in DIFMAP. As an alternative, we generated an L-band spectral index map in CASA and derived an average spectral index along with its upper and lower limits from a beam region. However, the S-band emission was too faint to extract an in-band spectral index during this analysis.

\section{Modeling Wide-band Radio Spectral Parameters}\label{app:spectra}
The observed radio spectra can be described by synchrotron mechanisms, where low-frequency absorption in the spectrum can be attributed to either free-free absorption (FFA) caused by ionized wind material along the line of sight or synchrotron self-absorption (SSA). A peaked radio spectrum can generally be modeled as \citep[see also][]{2022MNRAS.517.4959Y}
\begin{equation}\label{eq:sp}
F_\nu= \frac{f_p}{(1-e^{-1})}\times{\left(1-e^{{-(\nu/\nu_p)}^{\alpha_\mathrm{thin}-\alpha_\mathrm{thick}}}\right)}\times{\left(\frac{\nu}{\nu_p}\right)^{\alpha_\mathrm{thick}}},
\end{equation}
where $f_p$ is the flux density at the peak frequency $\nu_p$, and $\alpha_\mathrm{thick}$ and $\alpha_\mathrm{thin}$ are the spectral indices in the optically thick and optically thin regimes, respectively. In the framework of synchrotron emission driven by shock-CSM interaction, the radio evolution is approximately described by SSA \citep{1998ApJ...499..810C}, while FFA primarily steepens the light curves during the early phase. Therefore, we adopt the SSA mechanism here, with a spectral index of $\alpha_\mathrm{thick} = +2.5$, to model our spectra. Notably, the results of this work are independent of the specific absorption mechanism adopted. 
To fit the spectral parameters, we employed the Markov Chain Monte Carlo (MCMC) algorithm. The radio flux densities obtained on April 14, April 26, and June 26, 2022, were grouped as spectrum set I, while the simultaneous S- and C-band observations on September 11, 2023, formed spectrum set II. These two spectral sets represent the radio spectra at different epochs and were therefore fitted independently.

\begin{figure}
\centering
\includegraphics[width=0.47\textwidth]{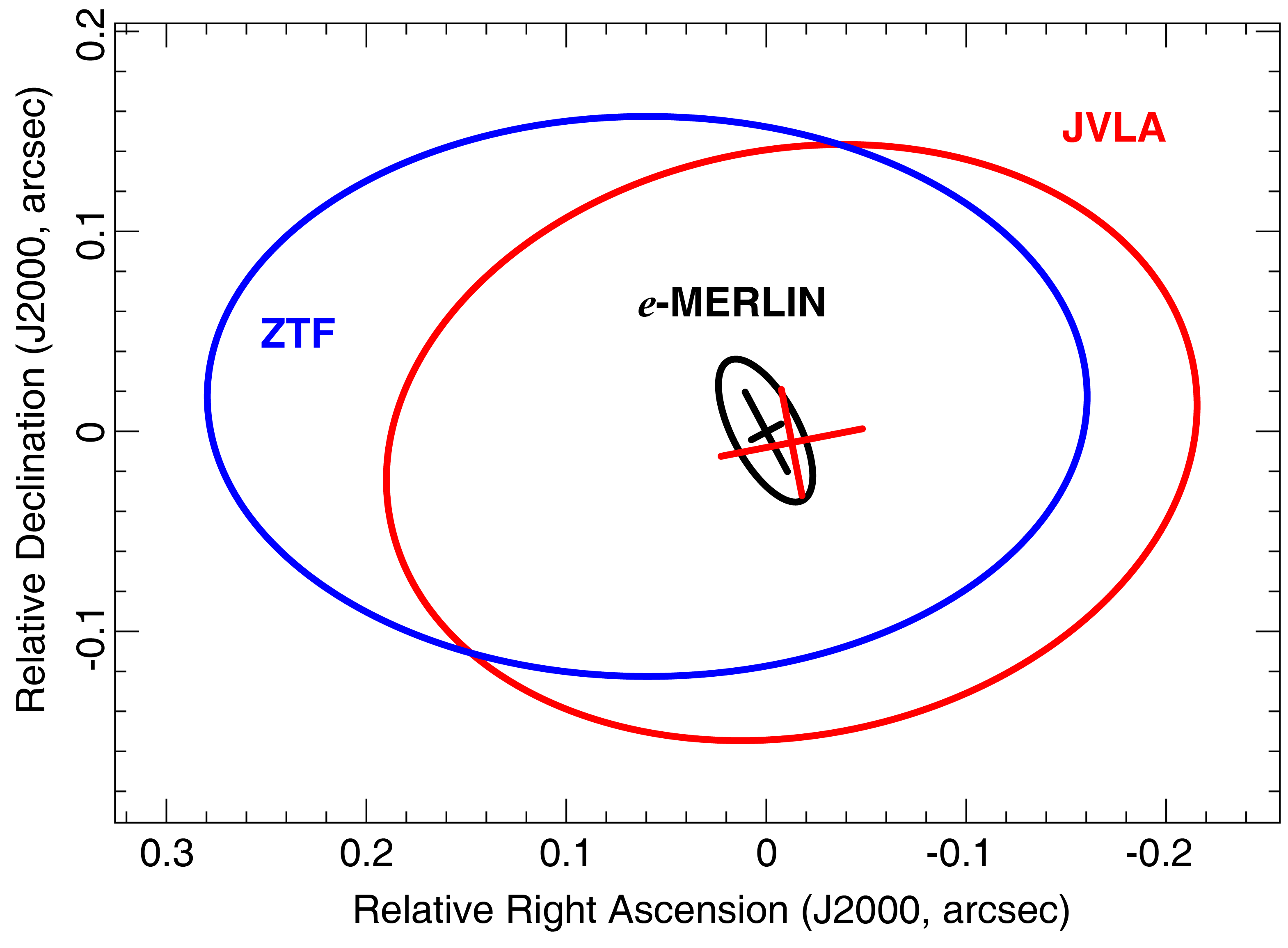}
\caption{Coincidence between radio and optical positions of \obj. The blue circle indicates the FWHM of optical position distributions estimated from ZTF, in the burst epochs. The red and black circles are the JVLA (project 22A-497 at C-band) and $e$-MERLIN (project RR12002) synthesize beam patterns, while the red and black crosses indicate the best positions and the corresponding $3\sigma$ uncertainties.}
\label{fig:astrometry}
\end{figure}

We explore the joint probability distribution of the free parameters (notably the peak frequency $\nu_{p1}$) for spectrum set I, while allowing variability only in the peak frequency $\nu_{p2}$ for spectrum set II and sharing the other parameters with spectrum set I. Our MCMC fitting yields a peak flux density of $f_p=101.89^{+9.49}_{-8.47}\,\mu\mathrm{Jy}$ at peak frequencies $\nu_{p1}=2.01^{+0.31}_{-0.28}\,\mathrm{GHz}$ and $\nu_{p2}=0.63^{+0.18}_{-0.16}\,\mathrm{GHz}$. The optically thin spectral index is $\alpha_\mathrm{thin} = -0.84^{+0.11}_{-0.12}$, corresponding to a power-law index of electrons, $p=2.68^{+0.24}_{-0.22}$. Interestingly, this is consistent with the assumption of $p\approx3$ in our modeling of the radio light curves. Assuming energy equipartition in the shocks (arising from the collision between the ejecta and CSM; \citealt{1998ApJ...499..810C}), we can estimate the outer shock radius accordingly, which is $2.79^{+0.31}_{-0.27} \times10^{16}$~cm and $8.95^{+2.56}_{-1.69}\times10^{16}$~cm from day 841 to day 1283 since the first detection, respectively, corresponding to a mean shock velocity of $17965^{+8278}_{-5860}$~km\,s$^{-1}$ over this period. However, this estimation is quite rough, primarily due to the lack of direct measurement of the spectral break in the second spectrum.

\begin{figure}
\centering
\includegraphics[width=0.47\textwidth]{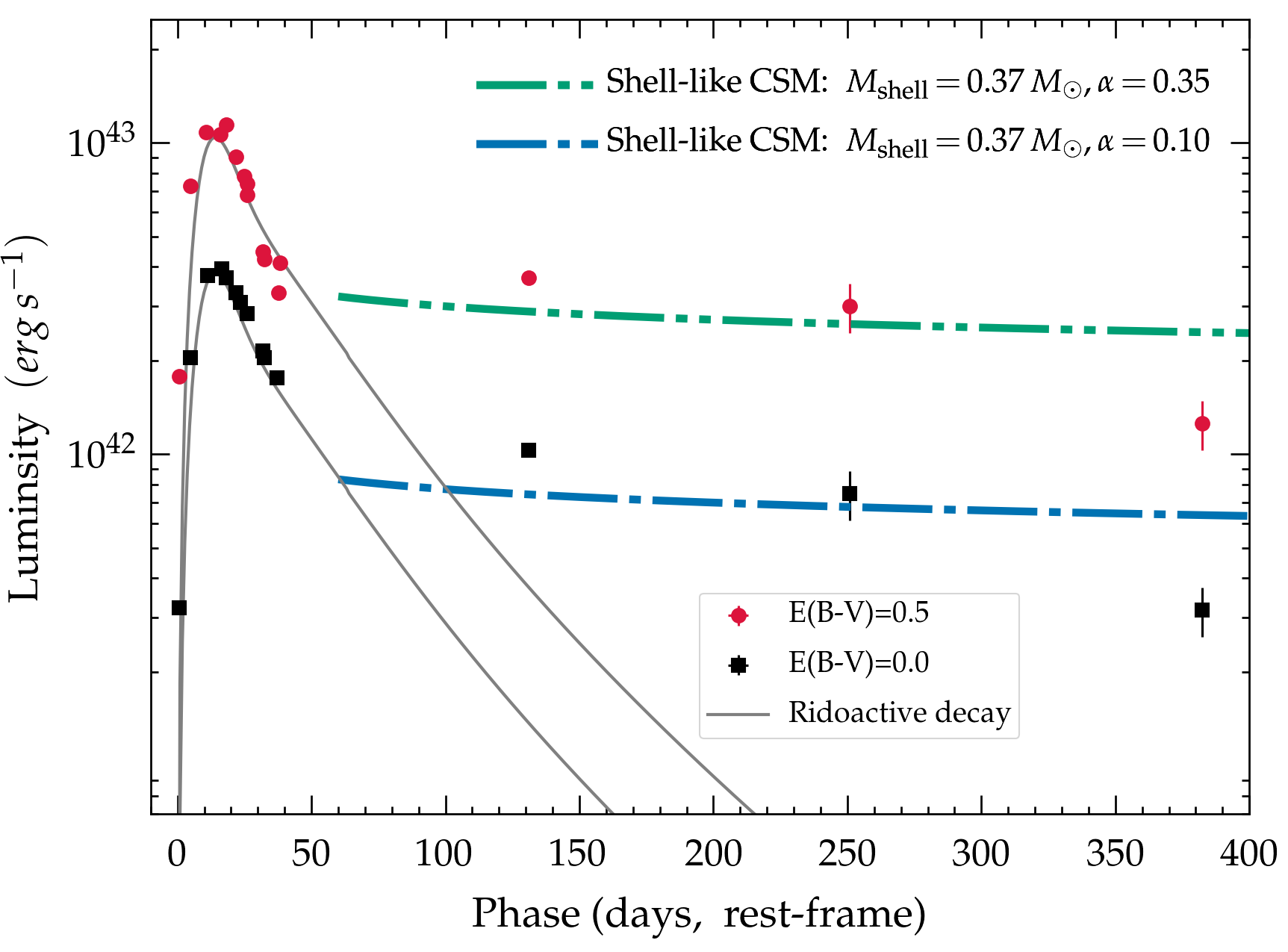}
\caption{Bolometric light-curve tails of SN~2020eyj fitted with the radioactive decay and the interaction between SN ejecta and a CSM-shell. The red circles and black squares present the bolometric light curves of SN~2020eyj for assuming an extinction of $E(B\,{-}\,V)\,{=}\,0.5$ and $E(B\,{-}\,V)\,{=}\,0.0$, respectively. The early peaks observed at ${\sim}20\,\mathrm{days}$ can be well reproduced by the radioactive decay of $\mathrm{^{56}Ni}$ (gray dotted lines).}\label{fig:shell_lc}
\end{figure}

\section{Astrometry of \obj\ and its Host Galaxy}\label{app:astrometry}
The Zwicky Transient Facility (ZTF\citealt{Smith2014,Bellm2019}) survey discovered SN~2020eyj on 7 March 2020 and monitored its decline over the subsequent $\sim400$ days until it faded into the background. We determine the location of SN~2020eyj detected by ZTF as the mean of coordinates reported in each ZTF $r$-band image between MJD 58915 and MJD~59200, during which SN~2020eyj was significantly brighter than the background and its host galaxy. Simultaneously, the position of the host galaxy was determined in the same way using ZTF $r$-band data before MJD~58915 and after MJD~59500, during which the optical luminosity was dominated by the host galaxy. Furthermore, we obtained optical coordinates of \obj\ reported by other alerts (Table~\ref{tab:astrometry}).

We obtain peak positions of radio emissions from $e$-MERLIN and JVLA observations as the radio coordinates of \obj\ (Table~\ref{tab:astrometry}). In our JVLA observations, the target SN~2020eyj was at the phase tracking center, and the phase correction of the target was referred to the nearby calibrator J1125$+$2610 every 6 min at both 6 and 10\,GHz. This phase referencing scheme can often assure an astrometric accuracy of $\sim10$\% of the synthesized beam FWHM \footnote{\url{https://science.nrao.edu/facilities/vla/docs/manuals/oss2023B/performance}}. The astrometric uncertainties in a phase referencing observation are generally caused by the noise of the phase calibrator and the thermal noise. The $1\sigma$ radio position uncertainty can be estimated as $\theta/(2\mathrm{SNR})$, where $\theta$ is the synthesized beam and $\mathrm{SNR}$ is the signal-to-noise ratio. In Figure~\ref{fig:astrometry}, we mark the best-determined radio coordinates from $e$-MERLIN (project code: RR12002) and JVLA C-band observations (project code: 22A-497), where $3\sigma$ positional uncertainties are taken.

\begin{deluxetable}{ccccccc}
\tablewidth{0pt}
\tabletypesize{\footnotesize}
\tablecaption{Astrometry of \obj\ and its host. \label{tab:astrometry}}
\tablehead{
\colhead{RA} & \colhead{DEC} & \colhead{Band} & \colhead{$\theta_\mathrm{maj}$} & \colhead{$\theta_\mathrm{min}$}   & \colhead{PA} & \colhead{Note} \\
\colhead{(hh:mm:ss)} & \colhead{(dd:mm:ss)}    & \colhead{} & \colhead{(arcsec)}       & \colhead{(arcsec)} & \colhead{(deg)} & \colhead{}
}
\colnumbers
\startdata
\multicolumn{7}{c}{Host} \\
\hline
11:11:47.160   &  +29:23:06.07   &    $r$    &$                 $&$                $&$         $&     PanSTARRS          \\
11:11:47.166   &  +29:23:06.18   &    $r$    &$    1.053        $&$  0.820         $&$   90    $&     ZTF (Quiescent state average) \\
\hline
\multicolumn{7}{c}{\obj\ Optical} \\
\hline
11:11:47.180   &  +29:23:06.36   &    G      &$                 $&$                $&$         $&    GaiaAlerts          \\
11:11:47.177   &  +29:23:06.46   &    $i$    &$                 $&$                $&$         $&    Pan-STARRS          \\
11:11:47.181   &  +29:23:06.44   &    $w$    &$                 $&$                $&$         $&    Pan-STARRS          \\
11:11:47.176   &  +29:23:06.48   &    $r$    &$                 $&$                $&$         $&    ZTF                 \\
11:11:47.180   &  +29:23:06.45   &    $r$    &$    0.440        $&$  0.280         $&$   90    $&    ZTF (Burst state average)\\
\hline
\multicolumn{7}{c}{\obj\ Radio} \\
\hline
11:11:47.176   &  +29:23:06.43   &    C      &$    0.079        $&$  0.033         $&$   28    $&    $e$-MERLIN (RR12002)  \\
11:11:47.177   &  +29:23:06.45   &    C      &$    0.093        $&$  0.071         $&$   107   $&    $e$-MERLIN (RR13005)  \\
11:11:47.177   &  +29:23:06.43   &    C      &$    1.016        $&$  0.283         $&$   61    $&    JVLA-A (22A-472)    \\
11:11:47.175   &  +29:23:06.43   &    C      &$    0.409        $&$  0.293         $&$   -79   $&    JVLA-A (22A-497)    \\
11:11:47.174   &  +29:23:06.43   &    X      &$    0.317        $&$  0.182         $&$   -73   $&    JVLA-A (22A-497)    \\
11:11:47.177   &  +29:23:06.43   &    C      &$    0.350        $&$  0.297         $&$   -73   $&    JVLA-A (23A-425)    \\
11:11:47.173   &  +29:23:06.38   &    C      &$    0.460        $&$  0.294         $&$   71    $&    JVLA-A (23B-299A)   \\
11:11:47.177   &  +29:23:06.38   &    C      &$    0.418        $&$  0.312         $&$   -69   $&    JVLA-A (23B-299C)   \\
11:11:47.173   &  +29:23:06.43   &    C      &$    0.491        $&$  0.299         $&$   71    $&    JVLA-A (23B-299D)   \\
\hline
\enddata
\tablecomments{Columns 1 and 2: Right Ascension and Declination (J2000); Column 3: observing band; Columns 4-6: major, minor, and position angle of FWHM of position distribution (optical) and synthesize beam (radio); Column 7: facility (observing code), names of surveys or all-sky alerts.}
\end{deluxetable}

\begin{deluxetable}{cccccc}
\tablewidth{0pt}
\tabletypesize{\footnotesize}
\tablecaption{The prior measured (before relative gain corrected) radio flux densities of in-beam calibrators. \label{tab:calibrator}}
\tablehead{
\colhead{Date} & \colhead{$\nu$} & \colhead{$f_\mathrm{p,A}$} & \colhead{$f_\mathrm{p,B}$} & \colhead{$f_\mathrm{p,C}$}   & \colhead{RMS} \\
\colhead{(UT)} & \colhead{(GHz)}    & \colhead{(mJy\,beam$^{-1}$)} & \colhead{(mJy\,beam$^{-1}$)}       & \colhead{(mJy\,beam$^{-1}$)} & \colhead{(mJy\,beam$^{-1}$)}
}
\colnumbers
\startdata
2022-04-14  &$  6     $&$   12.147          $&$   2.757             $&$    0.575               $&$   0.006      $ \\
2022-06-26  &$  6     $&$   11.391          $&$   2.674             $&$    0.558               $&$   0.004      $ \\
2023-07-11  &$  6     $&$   10.787          $&$   2.470             $&$    0.505               $&$   0.004      $ \\
2023-09-11  &$  6     $&$    9.654          $&$   2.122             $&$    0.438               $&$   0.004      $ \\
2023-09-29  &$  6     $&$   12.709          $&$   2.992             $&$    0.616               $&$   0.003      $ \\
2023-09-29  &$  6     $&$   11.945          $&$   2.635             $&$    0.531               $&$   0.004      $ \\
\hline
2022-06-26  &$  3     $&$   20.067          $&$   4.364             $&$    0.982               $&$   0.007      $ \\
2023-09-11  &$  3     $&$   15.110          $&$   3.186             $&$    0.745               $&$   0.007      $ \\
\hline
2022-04-26  &$  1.52  $&$   24.297          $&$   6.774             $&$    1.763               $&$   0.017      $ \\
2022-06-26  &$  10    $&$    4.427          $&$   1.381             $&$    0.225               $&$   0.003      $ \\
\hline
\enddata
\tablecomments{Column 1: observing date; Column 2: frequency; Columns 3-5: peak flux density of calibrators A, B, and C; Column 6: root-mean-square noise.}
\end{deluxetable}

\section{Modeling the bolometric light curve with a CSM shell}\label{app:lc}

 We use the analytical model of \citet{Liu2018} to calculate theoretical bolometric light curves for a spherically symmetric CSM-shell with a mass of ${\sim}\,0.37\,M_{\odot}$. We assume a power-law density profile for the outer SN ejecta, $\rho_{\rm ejecta} \propto r^{-n}$, with setting $n=7$. As shown in Fig.~\ref{fig:shell_lc}, the double-dot-dashed and dash-dotted lines represent the fits to the light-curve plateau of SN~2020eyj by assuming two different conversion efficiencies of $\alpha=0.35$ and $\alpha=0.1$, respectively, where $\alpha$ is an efficiency of converting the kinetic energy of SN ejecta to radiation during the CSM-shell interaction. Both models provide a reasonable reproduction of the observed bolometric light-curve plateau of SN~2020eyj, suggesting that interaction with a dense CSM shell can account for its sustained luminosity.

\clearpage
\bibliography{ref}{}
\bibliographystyle{aasjournalv7}

\end{document}